# Multiple superconducting phases in heavy-fermion metals

**Emilian M. Nica[1,*], Sheng Ran[2], Lin Jiao[3], Qimiao Si[4]**

[1]Department of Physics, Arizona State University, Tempe, AZ 85281, USA

[2]Department of Physics, Washington University in St. Louis, St. Louis, MO 63130, USA

[3]Center for Correlated Matter and Department of Physics, Zhejiang University, Hangzhou 310058, China

[4]Department of Physics and Astronomy, Rice Center for Quantum Materials, Rice University, Houston, TX 77005, USA

**\* Correspondence:**
enica@asu.edu

## Abstract

Symmetry breaking beyond a global U(1) phase is the key signature of unconventional superconductors. As prototypical strongly correlated materials, heavy-fermion metals provide ideal platforms for realizing unconventional superconductivity. In this article, we review heavy-fermion superconductivity, with a focus on those materials with multiple superconducting phases. In this context, we highlight the role of orbital-selective (matrix) pairing functions, which are defined as matrices in the space of effective orbital degrees of freedom such as electronic orbitals and sublattices as well as equivalent descriptions in terms of intra- and inter-band pairing components in the band basis. The role of quantum criticality and the associated strange-metal physics in the development of unconventional superconductivity is emphasized throughout. We discuss in some detail the recent experimental observations and theoretical perspectives in the illustrative cases of $UTe_2$, $CeRh_2As_2$, and $CeCu_2Si_2$, where applied magnetic fields or pressure induce a variety of superconducting phases. We close by providing a brief overview of overarching issues and implications for possible future directions.

## 1 Introduction

Unconventional superconductivity broadly refers to Cooper paired states which deviate from the Bardeen-Cooper-Schrieffer (BCS) paradigm, either in their pairing interactions or in the related issue of pairing symmetry (Bardeen, et al., 1957). It is a field with a history spanning more than four decades and several families of correlated materials. Heavy-fermion (HF) compounds enjoy a certain distinction: The first signatures of unconventional superconductivity were observed in $CeCu_2Si_2$ (Steglich et al., 1979), which together with fifty or so other subsequently discovered unconventional superconductors (SCs), belongs to this class (Stewart, 2017). A common feature of these materials is that the electron-electron interactions play a leading role in stabilizing a variety of low-temperature phases (Steglich and Wirth, 2016; Kirchner et al., 2020; Paschen and Si, 2021). Most of these HF compounds are either in the vicinity of or can be tuned near to quantum critical points (QCPs) where pronounced non-Fermi liquid behavior emerges. This can be traced to the fact that correlation effects, like Kondo screening, antiferro-/ferro-magnetic (AFM/FM), and valance fluctuations share similar small energy scales (Paschen and Si, 2021; White et al., 2015; Gegenwart et al., 2008; Coleman and Schofield, 2005; Weng et al., 2016). The interplay between these interaction-induced effects underlies the rich low-temperature physics.

## 1.1 Quantum criticality and HF superconductivity

This article is primarily concerned with the nature of the superconducting phases and, thus, we will be very brief in discussing the mechanism for HF superconductivity. Many HF superconductors develop at the border of magnetic order in their phase diagrams. Quantum criticality, and the associated strange-metal physics, is thus believed to cause superconductivity. Almost by definition, HF behavior involves Kondo screening, which is most naturally associated with Kondo-exchange interactions between quasi-localized $f$ and more dispersive conduction electrons. HFs exhibit characteristic scales associated with the initial onset of Kondo-singlet correlations, in the form of Kondo temperatures or coherence scales $T_{coh}$ (Steglich and Wirth, 2016; Kirchner et al. 2020). As already alluded to, the presence of several small scales in HF materials facilitates tuning in experiment, and is ultimately behind the emergence of rich phase diagrams. A hierarchy of multiple scales also leads to important general conclusions in the context of superconductivity. Indeed, based on the ratio of the critical superconducting and effective Fermi temperatures $T_c / T_F$, many HFs would qualify as high-temperature SCs (Steglich, 2014; Stewart, 2017). In addition, Ruderman-Kittel-Kasuya-Yoshida (RKKY) interactions between local moments, originating from quasi-localized $f$ electron states, are closely associated with Kondo interactions. RKKY and Kondo interactions typically compete, as proposed in the seminal work by Doniach (Doniach, 1977). It has been recognized in recent years that the RKKY-Kondo competition, together with the quantum fluctuations of the local moments, gives rise to a global phase diagram (Si 2006; Si 2010; Paschen and Si 2021). Extensive efforts, both experimental and theoretical, have been directed toward the elucidation of HF quantum criticality (Kirchner et al., 2020; Coleman et al., 2005; Löhneysen et al., 2007; Si and Steglich, 2010; Steglich and Wirth, 2016; Paschen and Si, 2021).

Broadly, two classes of QCPs have been advanced. One class is within the Landau framework where the quantum critical degrees of freedom (DOFs) correspond to the magnetic fluctuations (Hertz 1976; Millis 1993; Miyake et al., 1986; Mathur et al., 1998; Monthoux and Lonzarich, 2001). In the other, the Kondo effect itself is critical along with the fluctuations of the magnetic order parameter leading to Kondo-destruction (KD) QCPs (Si et al., 2001; Coleman et al., 2001; Senthil et al., 2004; Si et al., 2014; Steglich and Wirth, 2016; Kirchner et al., 2020; Paschen and Si, 2021). For the AFM case, it has been recognized that, even in the order-parameter fluctuation description i.e. nominally the first class, the quantum criticality is actually distinct from the original Hertz-Millis picture and is accordingly labeled SDW$_r$ quantum criticality instead (Hu et al. 2021a). Here, the subscript "r" marks the fact that the underlying quasiparticles are highly renormalized by the Kondo effect, which makes the Landau description valid only below a very small energy scale $k_B T_{cr}{}^*$. For scales above $k_B T_{cr}{}^*$ but below the bare Kondo scale $k_B T_K$, the quantum criticality also involves KD effects. Characteristic features of the KD quantum criticality include dynamical Planckian ($\frac{\hbar\omega}{k_B T}$) scaling and a sudden jump of "large" to "small" Fermi surface (FS) across the QCP. The amplified quantum fluctuations of the KD quantum criticality in the normal state have been shown to drive HF superconductivity (Hu et al. 2021a). Moreover, because of the involvement of the Kondo effect, the spectral weight of the quantum critical fluctuations is large. The resulting superconducting transition temperature $T_c$ is found to be high, reaching several percent of the effective Fermi energy (Hu et al., 2021a).

Experimental evidence for the KD quantum criticality has been extensively reported, especially in compounds such as $YbRh_2Si_2$ (Paschen et al., 2004; Friedemann et al., 2010; Prochaska et al., 2020), $CeCu_{6-x}Au_x$ (Löhneysen et al., 1994; Schröder et al., 2000), $CeRhIn_5$ (Shishido et al., 2005; Park et al., 2006; Knebel et al., 2008), and $Ce_3Pd_{20}Si_6$ (Custers et al., 2012; Martelli et al., 2019). In other cases, such as $CeCu_2Si_2$, there is evidence for quantum criticality in the form of magnetic order-parameter fluctuations (Arndt et al., 2011; Stockert et al., 2011, Smidman et al., 2018; Steglich and Wirth, 2016) and that, based on the temperature dependence of the spin damping rate and specific heat, $T^*_{cr}$ is about 1-2 K and is indeed small compared to the bare Kondo scale $T_K$ of about 20 K (Arndt et al., 2011; Gegenwart et al., 2008; Smidman et al., 2018; Smidman et al., 2022). Related effects on superconductivity, similarly due to quantum criticality are also expected in a number of cases, for example: $CeIn_3$ (Walker et al., 1997), $CeRh_2Si_2$ (Movshovich et al., 1996), $CePd_2Si_2$ (Grosche et al., 2001), and $CeIrSi_3$ (Sugitani et al., 2006). In this respect, HFs resemble other strongly-correlated materials like the cuprates, where superconductivity appears when AFM order, or the associated Mott insulator, is suppressed by chemical doping. Along the same lines, many Fe-based superconductors in the "1111" and "122" families also reach their highest $T_c$'s when the AFM order is suppressed. Moreover, non-magnetic quantum criticality in HF compounds, for instance associated with charge instabilities or multi-polar ordering as illustrated by examples covered previously, also appear to drive superconductivity.

We close this subsection with a remark on the energetics. For the heavy-fermion superconductors we will discuss, their normal states are typically at or close to being in the quantum critical regime. This means that their RKKY interaction is on the same order as their kinetic energy. As such, pairing between bands that are separated within the energy width of the RKKY-interaction-induced magnetic fluctuations will be energetically competitive.

## 1.2    Survey of HF systems with multiple superconducting phases

We now turn to reviewing HF superconducting regimes. Focusing in particular on HF systems with multiple superconducting phases, we show the ways in which the internal DOFs play an important role. A natural starting point is the case of superfluid $^3He$ (Leggett, 1975; Vollhardt and Wölfle, 1990). From a historic perspective, $^3He$ set the stage for Cooper pairing with nontrivial symmetry and (matrix) structure, both of which are associated with multiple superfluid phases. To illustrate, we consider the simplest pairing theories and the emergent conceptual framework. In a Ginzburg-Landau (GL) framework, the order parameter in $^3He$ is believed to preserve separate rotations in real space and spin space, as well as changes in the global phase. The relevant Cooper pairs have total angular momentum and spin 1, corresponding to odd-parity, spin-triplet pairing. The associated degeneracy is lifted in the Balian-Werthamer (BW) pairing state of the superfluid B phase, which is accessible at most pressures in zero magnetic field. In addition to selecting a global phase, the BW order parameter breaks a relative spin-orbit symmetry, by "locking together" the angular momentum and spin degrees of freedom (DOFs). It has a well-known $\hat{\mathbf{k}} \cdot (\sigma\, i\sigma_2)$ form which illustrates the residual combined $SO(3)_{L+S}$ rotation symmetry. The matrix structure which emerges as a consequence of the symmetry breaking is important for understanding the stability of this phase, and its most remarkable properties. Due to the emerging matrix structure in spin space, the BW phase is gapped, ensuring its predominant stability in the phase diagram, see Fig. 1(A). Moreover, the same structure leads to nontrivial topology and to the presence of Majorana edge modes (Nagato et al., 1998). In addition to the BW order parameter, a distinct Anderson-Brinkman-Morel (ABM) order parameter is associated with the superfluid A phase, which emerges in a restricted range of higher pressure and temperature in zero magnetic field. The ABM phase further breaks the symmetry by selecting preferred quantization axes for the total angular momentum and spin of the Cooper pair, respectively, with a residual $U_{L_z-\phi} \times U_{S_z}$ symmetry corresponding to rotations about the two axes combined with a change in the overall phase $\phi$. In its simplest realization, the ABM phase can be thought of as a $k_x + ik_y$ form factor multiplied by a trivial matrix corresponding to equal-spin pairing. The ABM order parameter also breaks time-reversal symmetry (TRS) and is one of the first instances of Weyl fermions in the context of condensed matter physics (Volovik, 2003). The phase diagram of $^3$He is more complex in the presence of a magnetic field, reflecting both the spin structure of the pairing, as well as the effective spin for the Cooper pairs associated with non-unitary order parameters (Leggett, 1975). For an exhaustive discussion we refer the reader to Refs. (Leggett, 1975; Vollhardt and Wölfle, 1990).

UBe$_{13}$ was the first instance of a solid-state system exhibiting multiple superconducting phases (Bucher, et al., 1975). Upon Th substitution at U sites, $U_{1-x}Th_xBe_{13}$ showed multiple superconducting phases driven by Th concentration or even temperature (Ott, et al., 1985). This discovery further supported the notion that superconductivity could be driven by unconventional pairing mechanisms. A more direct connection between superfluid $^3$He and unconventional superconductivity was provided by the HF compound UPt$_3$ (Stewart et al., 1984; Adenwalla et al., 1990). Indeed, at ambient pressure and in zero magnetic field, UPt$_3$ has two superconducting phases with decreasing temperature, see Fig. 1B (Huxley et al., 2002). In the presence of a field, this compound exhibits no less than five superconducting phases, among which are three flux phases, together with a tetracritical point (Joynt and Taillefer, 2002). While the phase diagram of UPt$_3$ suggests similarities with $^3$He, most notably spin-triplet pairing, the two systems also differ in ways which are shared by most HF SCs. In the presence of spin-orbit coupling (SOC), the Cooper pairs can be classified according to a symmetry group $G = D_{6h} \times \mathcal{T} \times U(1)$, corresponding to a finite point-group (PG) together with time-reversal and global phase rotations. The reduced rotational symmetry, as well as the presence of a space group appropriate to crystal structures, are hallmarks of solid-state superconductivity, in contrast to $^3$He. These effectively limit the pool of valid pairing candidates. Furthermore, the superconducting order parameter, as opposed to a superfluid, couples to electromagnetic fields, leading to additional superconducting regimes with applied fields. To describe the complex phase diagram of UPt$_3$, several nontrivial pairing candidates have been advanced. Since a comprehensive overview is prohibitive, here we touch upon those salient features which are in line with the presence of order parameters with nontrivial symmetry and matrix structure. In this context, we note that three of the most prominent pairing candidates, the so-called $E_{2u}$, various spin-triplet, and $E_{1g}$ pairing states belong to multi-dimensional irreducible representations of the PG (Joynt and Taillefer, 2002). Much like in $^3$He, the degeneracy inherent in these representations is lifted, or is conversely recovered, in the various superconducting phases. For instance, the two-dimensional, odd-parity, pseudo-spin-triplet $E_{2u}$ (Sauls, 1994) pairing state provides a minimal mechanism for the appearance of two superconducting phases at ambient pressure and in zero field. The $D_{6h}$ PG symmetry is broken due to the coupling of the order parameter to a basal-plane weak AFM order. The degeneracy of the two components belonging to the $E_{2u}$ irreducible representation is likewise lifted, resulting in two distinct transitions. The high-temperature phase selects one of the two components of $E_{2u}$. At lower temperatures, the superconducting order parameter regains a two-fold degeneracy in the form of a superposition of the two components with a $\pm \pi/2$ phase difference. Although the PG symmetry is preserved within a global phase rotation (Sauls, 1994), the system spontaneously breaks TRS. Remarkably, the two-component, even-parity, pseudo-spin singlet $E_{1g}$ pairing state has a similar predicted behavior, although it differs from $E_{2u}$ in the presence of finite magnetic fields (Joynt and Taillefer, 2002). While deviating to an extent from the order parameters of $^3$He, many of the pairing candidates in UPt$_3$ nonetheless reflect the intimate connection between symmetry-imposed degeneracy and the emergence of a variety of phases with distinct broken symmetries. In addition, the case of UPt$_3$ incorporates magnetic ordering and applied magnetic fields as crucial ingredients for the stability of multiple superconducting regimes. As one might expect, the behavior of the proposed pairing candidates exhibits considerable complexity in the presence of applied fields. We shall not discuss these important aspects here but instead refer the reader to Ref. (Joynt and Taillefer, 2002; Sauls, 1994) for a greatly expanded discussion on UPt$_3$.

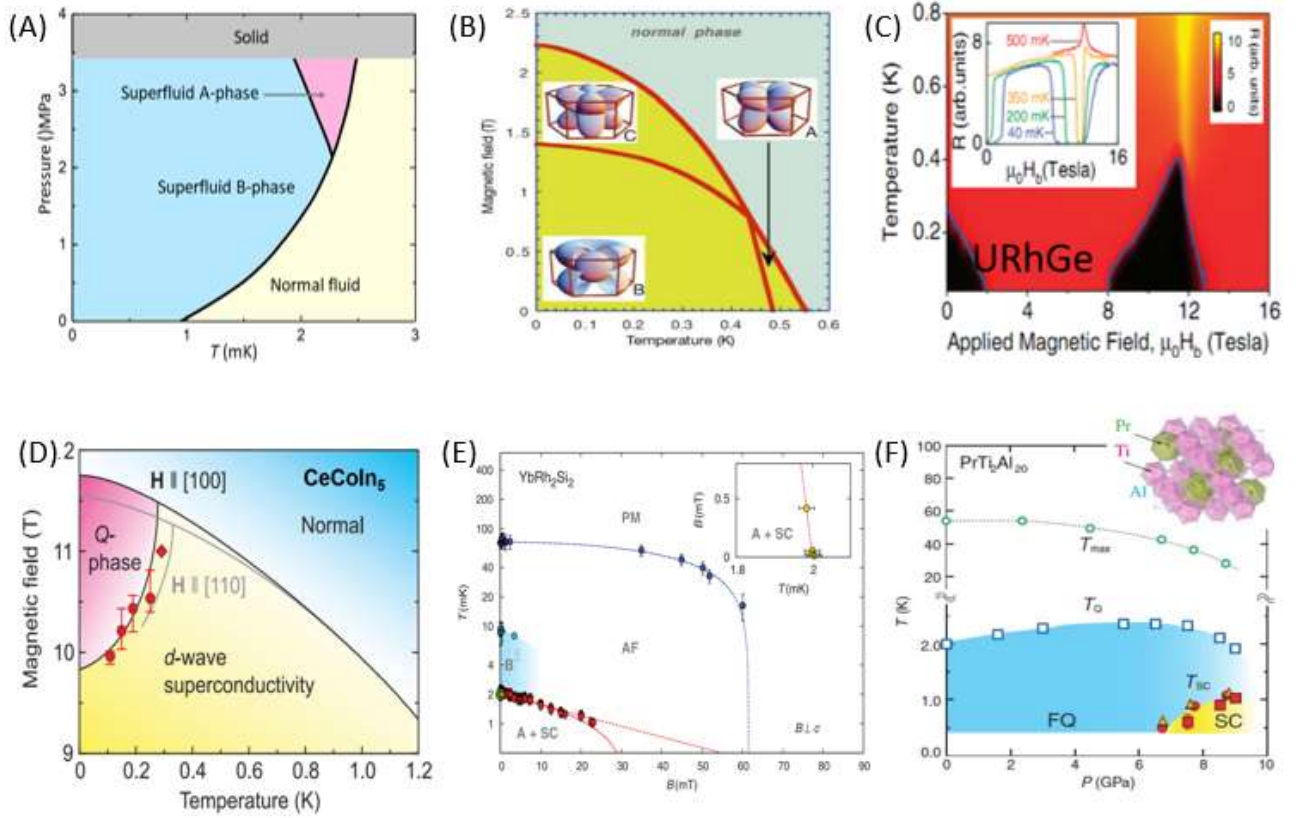

**Figure 1**. Low temperature phase diagrams of $^3$He (A), and of several heavy-fermion (HF) superconductors (SCs): (B) UPt$_3$ (Huxley et al., 2002), (C) URhGe (Aoki et al., 2001), (D) CeCoIn$_5$ (Kim et al., 2016), (E) YbRh$_2$Si$_2$ (Adapted from Schuberth et al., 2016; see also Nguyen et al., 2021), (F) PrTi$_2$Al$_{20}$ (Matsubayashi et al., 2012).

Before moving on to multiple superconducting phases in a number of prominent HF compounds, we further illustrate the extent of superconductivity in the broader HF family. Following the discovery of unconventional superconductivity in CeCu$_2$Si$_2$ (Steglich et al., 1979), several uranium-based superconductors have been subsequently identified, including the already-mentioned UBe$_{13}$ (Bucher et al., 1975), UPt$_3$ (Stewart et al., 1984), but also URu$_2$Si$_2$ (Palstra et al., 1985), prior to the appearance of the cuprates. Around the year 2000, instances of FM SCs, i.e., URhGe (Aoki et al., 2001), UGe$_2$ (Huxley et al., 2001), and UCoGe (Huy et al., 2007) have been reported. Unlike in Ce, where localized 4$f$-electrons underlie the Kondo effect, the 5$f$-electrons in U have more extended orbitals with higher itinerancy. In addition, 5$f$ electrons also have lower symmetry. In U-based compounds, 5$f$-electrons could also induce AFM, FM, and/or superconductivity. For example, Fig. 1C shows the phase diagram of URhGe (Aoki et al., 2001), where coexisting superconductivity and ferromagnetism have been observed. Since the magnetic moment is due to the itinerant 5$f$ electrons, the induced FM order has strong Ising anisotropy. The FM fluctuations favor an equal-spin pairing scenario, which in turn gives rise to strong anisotropic superconducting gap. More strikingly, a field-induced superconducting phase is observed in URhGe, which also implies a field-induced QCP in the vicinity of a metamagnetic transition.

Other HF compounds, such as CeCoIn$_5$ (Kim et al., 2016), YbRh$_2$Si$_2$ (Schuberth et al., 2016; Nguyen et al., 2021) and PrTi$_2$Al$_{20}$ (Matsubayashi et al., 2012) demonstrate rather unique superconducting regimes (Figs. 1D-F). A field-induced "Q-phase" is observed in CeCoIn$_5$, as shown in Fig. 1D. This phase was also proposed as a candidate to a Fulde-Ferrell-Larkin-Ovchinnikov (FFLO) state (Bianchi et al., 2003), in which the Cooper pairs have residual total momentum. Recent research indicated that this phase could be a pair density wave phase with a spatial modulation of the order parameter (Gerber et al., 2014). Similar superconducting phases have also been proposed in a few other compounds. Substituting Co with Rh or Ir generates the family of the "115" materials. Here, the interplay between superconductivity and antiferromagnetism has been thoroughly examined due to the availability of high-quality single crystals. Notably, a combined pressure and chemical doping study on Ce(Rh,Ir)In$_5$ reveals how a single superconducting dome splits into two. While the first dome is in the vicinity of an AFM QCP, the other emerges from the heavy-Fermi liquid state without any trace of critical magnetic fluctuations, suggesting that charge/valence instabilities play leading roles instead (Kawasaki et al., 2006). Nevertheless, this picture needs further clarifications as evidence for AFM fluctuation has been observed for pure CeIrIn$_5$ at ambient pressure by Shang et al. (Shang et al., 2014).

The compound YbRh$_2$Si$_2$ is a prototypical HF exhibiting an AFM QCP (Trovarelli et al., 2000; Gegenwart et al., 2002; Custers et al., 2003; Ernst et al., 2011). It has become a canonical case of KD QCP (Si et al, 2001; Paschen et al., 2004; Prochaska et al., 2020) and been discussed extensively as such (Steglich and Wirth, 2016; Kirchner et al., 2020; Paschen and Si, 2021). However, superconductivity in this compound was not detected until recently. Upon being cooled down to the mK range, superconductivity was observed in YbRh$_2$Si$_2$ using magnetic susceptibility (Schuberth et al., 2016) and electrical resistivity measurements (Nyuyen et al., 2021). In Ref. (Schuberth et al., 2016), the development of a hybrid nuclear-electronic magnetic order is also observed. It was proposed that the nuclear-spin ordering reduces the primary electronic order, thereby pushing the system closer to the KD quantum criticality, which promotes unconventional superconductivity (Schuberth et al., 2016). Upon replacing the naturally abundant Yb by $^{174}$Yb, which has no nuclear magnetic moment, superconductivity is weakened (Nguyen et al., 2021), a result that is consistent with the role played by the nuclear spins. Nguyen et al. furthermore provided evidence for two different superconducting phases in YbRh$_2$Si$_2$ that, in the $^{174}$Yb case, splits into two adjacent domes, thus making YbRh$_2$Si$_2$ a new example of a multiple-phase SC (Nguyen et al., 2021). This is an exciting result, suggesting that the pair-breaking effect of the external magnetic field is not as severe as previously thought, and that the innate quantum criticality at the quantum critical field can still nucleate superconductivity (Nguyen et al., 2021). Taken together, these experiments (Schuberth et al., 2016; Nyuyen et al., 2021) provide compelling evidence that superconductivity is driven by the KD quantum criticality. Moreover, the high-field superconducting phase was tentatively identified as a potential case of spin-triplet pairing (Nguyen et al., 2021), a result that is supported by the calculations of pairing correlations near the KD QCP of a cluster Bose-Fermi Kondo model in the Ising-anisotropic case (Pixley et al., 2015; Hu et al., 2021b), especially when the Zeeman effect of the external magnetic field is considered (Hu et al., 2021b).

Another example of complex superconducting order in HF is provided by PrTi$_2$Al$_{20}$ (Sakai et al., 2012; Matsubayashi et al., 2012). Just like a local magnetic moment can be screened by itinerant electrons in a conventional HF compound, the electric quadrupole moment of Pr$^{3+}$ can in principle be screened by conduction electrons. The $f$-electrons in PrTi$_2$Al$_{20}$ belong to low energy, non-magnetic (non-Kramers) $\Gamma_3$ doublets which further transition into a long-range ferro-quadrupole (FQ) ordered state at low temperatures. Superconductivity appears upon suppressing the FQ order via high pressure. Around 9 GPa, the superconducting transition temperature and the effective masses of the heavy quasiparticles are dramatically enhanced, indicating the presence of a QCP associated with quadrupolar order.

In spite of the multitude of superconducting phases alluded to previously, virtually all HF compounds are expected to share a number of common features, which can also guide our understanding of the pairing mechanisms. We have already mentioned the Kondo-RKKY competition and the resulting quantum criticality in the normal state. As also alluded to earlier, the presence of several small scales in HF materials facilitates tuning in experiment, and is ultimately behind the emergence of rich phase diagrams. We refer to Sec. 1.1 for further discussions in this regard.

## 1.3 Orbital-selective (matrix) pairing states

According to the BCS theory, the formation of Cooper pairs spontaneously breaks the U(1) gauge symmetry. Unconventional SCs can further break crystal symmetry (rotation or inversion) and/or TRS. The latter often involves the effective release of additional DOFs, such as spin, orbital, quadrupole, valley et al. Given the success of GL theory within the canonical BCS paradigm, and in the case of superfluid $^3$He (Leggett 1975; Vollhardt and Wölfle, 1990), it is not altogether surprising that most studies of unconventional superconductivity in complex, multiband HF compounds have focused on a broad, symmetry-based classification of the pairing states according to the PG. For spin-singlet pairing, the central question typically being asked is whether the pairing state is an (effectively) single band $s$- or $d$-wave, with an implicit understanding that the two cases are mutually exclusive. Similar symmetry-based classifications approaches have been adopted for spin-triplet superconductors (Joynt and Taillefer, 2002; Sigrist and Ueda, 1991).

One focus of our discussion concerns the nature of the pairing states when there are multiple internal DOFs are involved, such as orbitals and sublattices. Such an extension of the set of candidate pairing states, which goes beyond the effectively single-band picture and which has immediate implications for HF superconductivity, has been driven by recent surprising experimental results. A notable case in this respect is that of the alkaline Fe-selenide compounds (Si et al., 2016; Lee 2017). Here, multiple ARPES experiments detected fully-gapped superconductivity (Wang et al., 2011; Xu et al., 2012), normally associated with a fully-gapped $s$-wave pairing candidate. By contrast, inelastic neutron scattering (INS) experiments detected the presence of a resonance in the spin spectrum (Park et al., 2011; Friemel et al., 2012), which requires a sign-change in the order parameter between different FS sheets. Due to the presence of a small hole pocket near the center of the Brillouin Zone (BZ) (Wang et al., 2011; Xu et al., 2012), the INS experiments would suggest a $d$-wave pairing candidate instead. As discussed in greater detail below, a remarkably similar experimental picture subsequently emerged in the HF CeCu$_2$Si$_2$.

To address the discrepancy between the experiments, which suggest in turn incompatible $s$- and $d$-wave single-band pairing candidates, Ref. (Nica et al., 2017) proposed an $s\tau_3$ pairing candidate by taking into account the multi-orbital/-band nature of the alkaline Fe-selenides. In addition to the usual spin-singlet matrix form, this orbital-selective pairing also incorporates an additional $\tau_3$ matrix form in orbital space, defined for the most relevant $d_{xz/yz}$ orbitals. The $\tau_3$ matrix structure indicates that the pairing has opposite phases in the two orbital sectors, respectively. The additional "s" stands for a standard $s$-wave form factor. It is crucial to note that $s\tau_3$ still belongs to a single irreducible representation of the $D_{4h}$ PG, much like $s$- and $d$-wave effective single-band candidates. As we shall see in subsequent sections, pairing states which incorporate additional (effective) orbital structure much like the $s\tau_3$ candidate have also been proposed for several distinct HFs that involve multiple SC phases.

In the alkaline Fe-selenides, the $s\tau_3$ pairing function does not commute with the kinetic part of the Hamiltonian, thus ensuring that the pairing is equivalent to two simultaneous but distinct *intra-* and *inter-band* d-wave components. The corresponding matrix pairing in band space can thus be classified as $d+d$ pairing. Precisely due to the matrix structure, the two d-wave components add in quadrature to produce a full gap. Furthermore, they change sign under a $\pi/2$ rotation. These two aspects ensure that $s\tau_3$ $(d+d)$ pairing can reconcile the experiments in the alkaline Fe-selenides. As for the experimental results, $d+d$ pairing can be extended to HF superconductivity, as discussed in greater detail below.

The matrix structure of $d + d$ pairing in the band basis bears a resemblance to the matrix structure of superfluid $^3$He-B, defined in a spin basis (Nica et al., 2021), which hints at a natural generalization of matrix pairing. While the $d + d$ pairing state mixes two distinct d-wave components via a nontrivial matrix structure in band space, the canonical pairing state for $^3$He-B mixes three distinct p-wave components in a matrix structure which combines equal- and opposite-spin fermions, with the consequence that the FS is gapped in both cases. Moreover, states with a trivial matrix structure in band space, i.e., proportional to an identity matrix, such as a $d + id$ pairing candidates, are conceptually similar to the most basic equal-spin pairing in $^3$He-A. Thus, $^3$He in general provides important precedents for nontrivial matrix pairing states. It can guide our understanding of matrix pairing states generalized to orbital/spin-orbital DOF in crystalline SCs, in spite of the differences in parity, dimensionality, symmetry, and symmetry-breaking.

An important feature of the Fe-chalcogenide (as well as Fe-arsenide) superconductors is that the band width of the magnetic fluctuations is large, typically on the order of 200 meV (Dai, 2015). Such a large bandwith is associated with the large magnitude of the short-range spin-exchange interactions. As a result, the involved interband pairing is expected to be energetically competitive. Indeed, for microscopic models of the alkaline Fe-selenides, the $s\tau_3$ pairing state has been shown to be energetically competitive (Nica et al., 2017).

In the following, we review the recent progress in understanding multiple superconducting phases in HFs, as exemplified by UTe$_2$, CeRh$_2$As$_2$, and CeCu$_2$Si$_2$. We illustrate the remarkable tunability of superconductivity in these compounds and discuss evidence in favor of pairing states with nontrivial symmetry and structure.

## 2    Multiple superconducting phases in heavy-fermion superconductors

### 2.1    Pressure and magnetic field induced multiple superconducting phases in UTe$_2$

UTe$_2$ is a recently discovered spin-triplet candidate SC (Ran et al., 2019). While the superconducting state of UTe$_2$ closely resembles that of FM SCs, e.g., URhGe (Aoki et al., 2001) and UCoGe (Huy et al., 2007), the normal state of UTe$_2$ is paramagnetic (Ran et al., 2019a; Aoki et al., 2019). Nonetheless, the pronounced magnetic susceptibility (along one crystalline direction) and the H/T$^{1.5}$ scaling of the magnetization provided initial indication for the importance of FM quantum fluctuations (Ran et al., 2019a). Spin-triplet pairing is strongly indicated by the extremely large, anisotropic upper critical field $H_{c2}$, nodes on the superconducting gap, and the temperature-independent NMR Knight shift in the superconducting state (Ran et al., 2019a; Aoki et al., 2019; Metz et al., 2019; Bae et al., 2021; Nakamine et al., 2019; Nakamine et al., 2021). A nontrivial topology is also suggested by the observation of chiral in-gap bound states from scanning tunneling spectroscopy (Jiao et al., 2020). As expected for unconventional superconductors, UTe$_2$ likely has a nontrivial order parameter. It hosts multiple superconducting phases when magnetic field or external pressure is applied, some of which hold surprisingly high temperature and magnetic field stability. In the scope of this brief review, we focus on the order parameter and on the multiple superconducting phases of UTe$_2$.

The initial specific heat measurements showed a single transition at $T_c$. However, when the specific heat was measured by means of the small-pulse method, a shoulder-like feature appears at a temperature of about 75 to 100 mK above the peak in $C_p/T$ (Hayes et al., 2021). This feature is quite sharp and divides the jump in the specific heat into two local maxima in the derivative, d($C_p/T)$/d$T$, representing two thermodynamic anomalies. In principle, the two transitions could be the result of inhomogeneity in the crystals. On the other hand, if there is indeed a multicomponent order parameter, TRS could be broken, which cannot be explained by inhomogeneity. To test for possible TRS breaking in the superconducting state of UTe$_2$, high-resolution polar Kerr effects measurements were performed using a zero-area Sagnac interferometer (Hayes et al., 2021). Initial measurements indicated that without an applied magnetic field, the Kerr signal can be either finite or zero. This observation led to the speculation that TRS breaking domains form spontaneously upon cooling the sample, which can orient in opposite directions and give average signal of zero.

To orient all of the domains in one direction, a small, field of +25 G was applied upon cooling the sample, and removed once the base temperature (~300 mK) was reached. The Kerr angle was subsequently measured as the sample was warmed slowly, and it was found that a positive finite Kerr value develops around $T_c$ in this zero-field measurement. The sign of the Kerr value was reversed with a negative training field (−25 G), indicating that the broken TRS order parameter is analogous to a magnetic moment (Hayes et al., 2021), as shown in Fig. 2. As a finite Kerr signal cannot be explained by the inhomogeneity, the combination of the specific heat and Kerr effect measurements point to a multi-component order parameter.

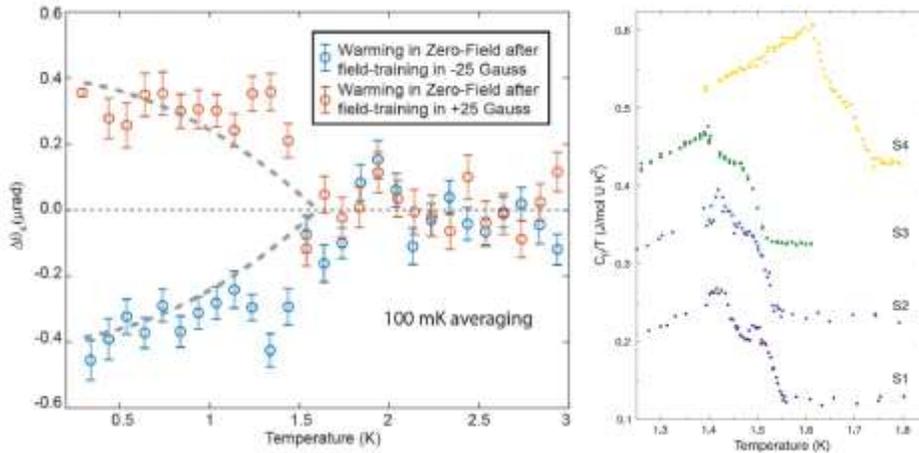

**Figure 2**. **a**. Finite Kerr signal below the superconducting transition temperature of UTe$_2$ indicating time reversal symmetry breaking. **b**. Specific heat of UTe$_2$ showing should-like feature at critical temperature. After (Hayes et al., 2021).

### 2.1.1 Re-entrant superconducting phases in applied magnetic fields

In applied magnetic fields, two independent, high-field superconducting phases were discovered in UTe$_2$ (Ran et al., 2019b; Knafo et al., 2021), for a total of three superconducting phases, as shown in Fig. 3. This is an example of two field-induced superconducting phases emerging in a single compound, one of which has the highest lower and upper limiting fields of any field-induced superconducting phase: more than 40 T and 65 T, respectively.

When the magnetic field is perfectly aligned along the $b$-axis, zero resistance persists up to 34.5 T at 0.35 K, with a transition from $SC_{PM}$ to $SC_{RE}$ at around 21 T. While evidence from thermodynamic measurements is absent, the transition between two separate superconducting phases is clear when the magnetic field is slightly rotated away from $b$ axis, towards either $a$- or $c$-axis, as shown in Fig. 3. For example, a small misalignment of less than 5° from the $b$-axis towards the $a$-axis decreases the $H_{c2}$ value of $SC_{PM}$ to 15.8 T. When the magnetic field is further increased, a field-induced superconducting phase $SC_{RE}$ appears between 21 T and 30 T. Resistance measurements show that this re-entrant phase, $SC_{RE}$, is present in a small angle range, within 7° when the field is rotated from $b$-axis towards the $a$-axis, and within 4° towards the $c$-axis (Ran et al., 2019b).

The upper-field limit of $SC_{RE}$ of 35 T coincides with a dramatic magnetic transition into a field-polarized phase, evidenced by a magnetic moment change from 0.35 to 0.65 $\mu_B$. The critical field $H_m$ of this magnetic transition has little temperature dependence up to 10 K, but $H_m$ increases as the magnetic field rotates away from the $b$-axis to either the $a$- or $c$-axis (Ran et al., 2019; Knafo et al., 2021; Miyake et al., 2019). As $H_m$ limits the $SC_{RE}$ phase, it gives rise to the second field-induced superconducting phase, $SC_{FP}$, existing in the angular range of $\theta = 20$–40° from the $b$-axis towards the $c$-axis. The onset field of the $SC_{FP}$ phase precisely follows the angle dependence of $H_m$, while the upper critical field goes through a dome, with the maximum value exceeding 65 T, the largest field strength of these measurements. With $T_c$ of 1.5 K, this new superconducting phase largely exceeds the magnetic field range of all known field-induced superconducting phases. Its special angle range is roughly the normal direction of the (011) plane, the easy cleave plane of $UTe_2$, indicating a quasi-2D nature for this superconducting phase. These results are summarized in Fig. 3.

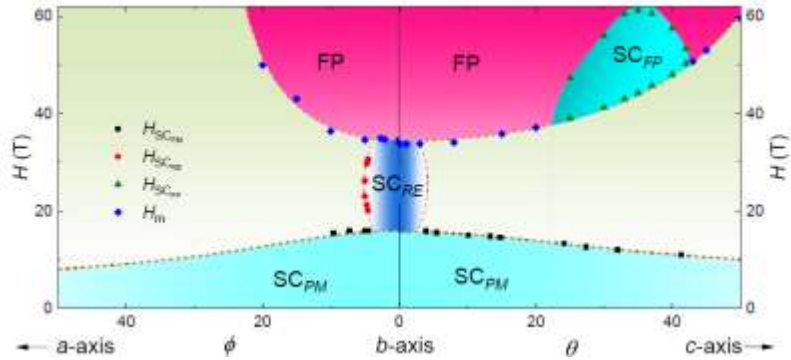

**Figure 3**. Magnetic field–angle phase diagram of $UTe_2$ showing the three superconducting phases $SC_{PM}$, $SC_{RE}$ and $SC_{FP}$. After (Ran et al., 2019).

### 2.1.2 Multiple superconducting phases under applied pressure

The critical temperature $T_c$ of $UTe_2$ shows a two-fold enhancement under pressure (Braithwaite et al., 2019; Ran et al., 2020; Thomas et al., 2020). Below 1.31 GPa, $T_c$ forms a clear dome feature under pressure peaked at 1 GPa, where $T_c$ is doubled compared to the ambient-pressure value, reaching 3.2 K. The bulk nature of the SC is confirmed by magnetization data. Specific heat measurements indicate a phase transition from an ambient-pressure superconducting phase to a SC2 phase under higher pressure (Aoki et al., 2020), as shown in Fig. 4. A new phase, magnetic in nature, appears above 1.5 GPa, which has been assigned to both FM and AFM order (Ran et al., 2020; Thomas et al., 2020). On one hand, hysteresis has been observed, indicating a possible FM nature. On the other hand, the critical temperature of the magnetic phase is suppressed with applied field in all directions, suggesting an AFM order instead. The shapes of the resistivity curves at the phase transition are also more consistent with antiferromagnetism. The exact nature of this phase needs to be confirmed by additional experiments.

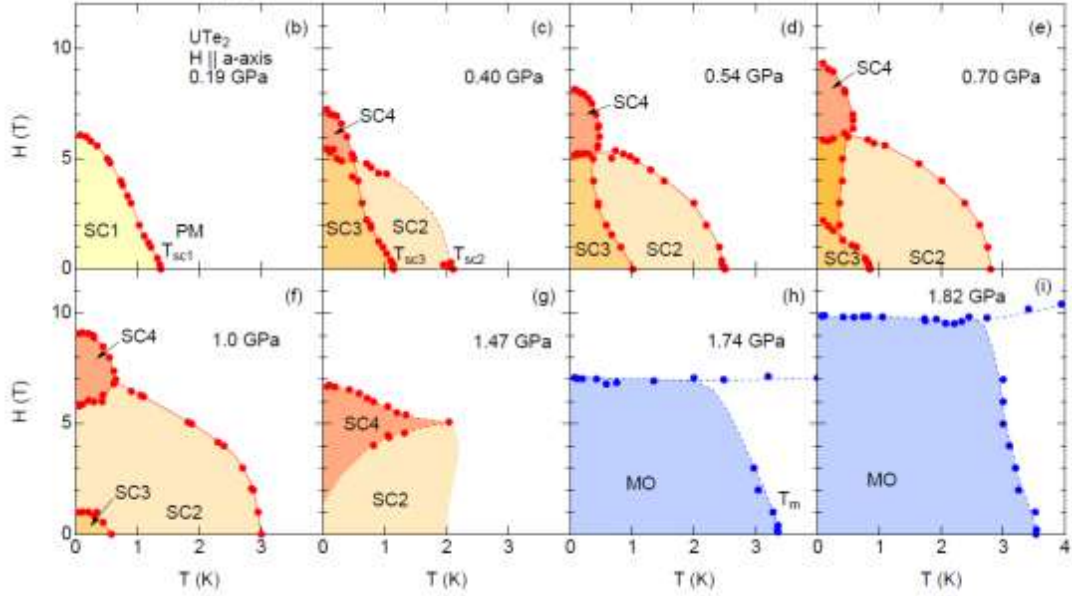

**Figure 4**. *H-T* phase diagrams of UTe$_2$ for *H // a*-axis at several pressures. After (Aoki et al., 2020).

A remarkably rich phase diagram emerges when the magnetic field is further applied under pressure. With (externally-applied) pressure, the magnetic field cannot be rotated relative to the sample *in-situ*. Therefore, only a few magnetic field directions have been explored: along the *a*-axis, *b*-axis, and for those special angles where high field induced SC$_{FP}$ appears.

Specific heat and resistance measurements have been carried out under pressure for magnetic fields applied along the *a*-axis (Fig. 4). Within the zero-resistance region, different features in the specific heat curves have been used to delineate phase boundaries between different superconducting regimes. In the low-pressure region, a single superconducting phase was observed under magnetic field. As pressure was increased, SC1 disappeared, in agreement with the zero magnetic field results, and up to three new superconducting phases were observed. One phase only exists in the applied magnetic field along *a*-axis. Another emerged only at higher temperatures. The phase diagram in the intermediate pressure region, ~0.5 GPa, resembles that of the UPt$_3$ at ambient pressure (Joynt and Taillefer, 2002). Given the important role of specific heat measurements in determining superconducting phase boundaries, it is conceivable that similar measurements might reveal additional paired phases when the magnetic field is applied in other directions. However, such measurements have not yet been performed. Discussions for other field directions are based mainly on transport measurements.

As already mentioned, when the field was applied along the *b*-axis, two superconducting phases, SC$_{PM}$ and SC$_{RE}$ were observed, as shown in Fig. 5 (A). SC$_{RE}$ persisted up to the metamagnetic phase transition at $H_m$. Under applied pressure, $H_m$ remained close to the upper limit of SC$_{RE}$ and was continuously suppressed, even though the critical temperature of SC$_{RE}$ was enhanced (Lin et al., 2020; Ran et al., 2021). Based on tunneling diode oscillator and resistance measurements, the phase boundaries between $SC_{PM}$ and $SC_{RE}$ were determined, which were also suppressed under pressure. To study the evolution of the high-field superconducting phase, $SC_{FP}$, the magnetic field was applied 25°-30° away from the *b* axis toward *c* (Fig. 5 (B)). At ambient pressure, two superconducting phases, $SC_{PM}$ and $SC_{FP}$, were established. Upon an initial increase in pressure, the stability of both superconducting phases was enhanced: the upper critical field of $SC_{PM}$, $H_{c2}$, increased, and the critical onset field of $SC_{FP}$, which coincided with the metamagnetic transition field, decreased. In an intermediate crossover pressure range, the phase boundary between $SC_{PM}$ and $SC_{FP}$ was no longer visible in the electrical resistance; resistance remained zero up to 45 T at base temperature, which was the largest DC magnetic field available to the experiment (Ran et al., 2021).

Under pressure, the metamagnetic phase transition remains pinned to the upper limit of $SC_{PM}$ phase. As the pressure further increases, the metamagnetic transition is suppressed, as is the upper critical field of $SC_{PM}$. But the critical onset field of $SC_{FP}$ starts to increase, and the two superconducting phases are no longer connected above 1.2 GPa. When the metamagnetic transition vanishes under a pressure of 1.4 GPa, $SC_{PM}$ is also suppressed completely. $SC_{FP}$ persists under the highest pressure in these studies. The low- and high-field superconducting phases always exist on opposite sides of the metamagnetic transition HFP, for which a FS reconstruction scenario has been suggested based on thermoelectric power and Hall effect measurements. These imply that the $SC_{PM}$ and $SC_{FP}$ phases which are separated by the metamagnetic transition might exhibit different pairings states that are unique to PM and FP phases, respectively.

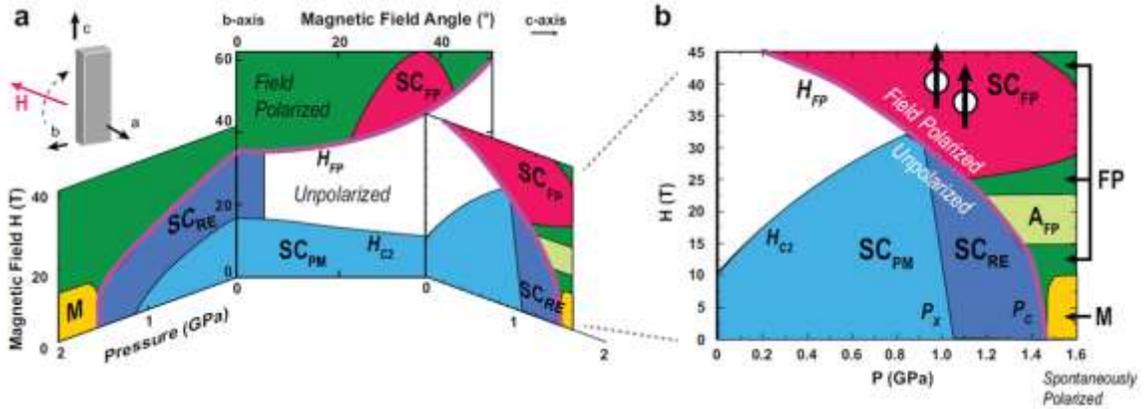

**Figure 5**. **(A)**. *H-P* phase diagram of UTe$_2$ for magnetic fields applied along *b*-axis. (B) *H-P* phase diagram of UTe$_2$ for magnetic fields applied 25°-30° away from the *b*-axis toward *c*-axis. After (Ran et al., 2021).

### 2.1.3 Theoretical perspective

The novelty of superconductivity in UTe$_2$ and the complexity of the phase diagram, as a function temperature, pressure, and magnetic fields, preclude a firm consensus on the pairing candidates at this point (Aoki et al., 2022). We will not attempt a comprehensive review of existing theories but will instead briefly cover a few representative studies in an attempt to identify those points which appear to distinguish UTe$_2$ from other unconventional superconductors. Given the strong evidence for spin-triplet pairing, many proposals focus on this case. Even so, two major issues tentatively set UTe$_2$ apart. First, the orthorhombic lattice and the associated $D_{2h}$ PG symmetry restrict the order parameter to one-dimensional irreducible representations. The degeneracy of the order parameters in other spin-triplet candidates, like $^3$He and UPt$_3$, which can naturally account for the presence of multiple broken-symmetry states, does not seem to hold for UTe$_2$. It should be mentioned that proposals based on accidental degeneracy between one-dimensional irreducible representations, which appear to naturally account for the observation of two superconducting transitions in UTe$_2$ in zero field and ambient pressure, have also been advanced for UPt$_3$ (Joynt and Taillefer, 2002). Secondly, the absence of FM fluctuations and ordering, as determined by inelastic neutron scattering (INS) measurements (Duan et al., 2021), which are expected to stabilize spin-triplet pairing, provides a challenge for theory. Indeed, INS experiments rather point to dominant AFM fluctuations instead. In this regard, UTe$_2$ can also be contrasted to UPt$_3$, where a weak AFM order plays an important role in lifting the degeneracy inherent to multi-dimensional $E_{1g}$ and $E_{2u}$ irreducible representations (Sauls, 1994; Sigrist and Ueda, 1991; Vollhardt and Wölfle, 1990). It is tempting to note that AFM fluctuations in UTe$_2$, as evidenced by INS experiments, and in UPt$_3$, where an incipient AFM order was established, appear to play important roles for both candidates to spin-triplet pairing. Although not conclusive at this stage, this commonality suggests that AFM critical fluctuations, in the presence of anisotropy inherent to materials with significant SOC, might also provide a pairing mechanism for spin-triplet SCs.

We note that pairing candidates typically take into account the band structure and FS, the low-energy gapped or gapless Bogoliubov-de Gennes (BdG) spectrum, the pairing mechanism, the existence of multiple superconducting phases, and nontrivial topology. In the first category, we note that a number of *ab initio* studies predict a similar shape of the FS (Ishizuka et al., 2019; Xu et al., 2019; Nevidomski 2020; Miao et al., 2020; Duan et al., 2020), consisting of quasi-2D cylinder-like electron and hole FS sheets, although other shapes were also advanced (Shishidou et al., 2021; Miao et al., 2020). In a related manner, a number of authors proposed non-unitary, mixed-representation, TRS-breaking spin-triplet states (Ran et al., 2019; Aoki et al., 2019; Hayes et al., Science 2021) such as $B_{1u} + iB_{3u}$ (Nevidomski, 2020) or $B_{3u} + iB_{2u}$ (Shishidou et al., 2021) which can in principle account for the presence of point nodes and topologically nontrivial chiral edge states consistent with experiments. Possible nontrivial topology is further discussed in Refs. (Ishizuka et al., 2019, Shishidou et al., 2021). As noted previously, the quasi-degeneracy of the two components included in these mixed representation states is not guaranteed by symmetry. Nonetheless, such candidates can in principle also provide a resolution to the presence of multiple superconducting phases with applied fields, especially since non-unitary pairing states exhibit nontrivial effective magnetic moments which can couple to magnetism, like in UPt$_3$, at least at the conceptual level. Note that TRS-preserving unitary states have also been advanced (Xu et al., 2019; Ishizuka et al., 2019; Ishizuka and Yanase, 2021). Finally, the interplay between AFM and FM fluctuations leading to triplet pairing has been addressed by a number of recent works (Hu and Si, 2020; Chen et al., 2021; Kreisel et al., 2022).

FM, *intra-unit cell* exchange interactions have been especially invoked as promoting triplet pairing (Hu and Si, 2020; Shishidou et al., 2021; Chen et al., 2021). This has led to a matrix pairing state via additional effective orbital DOF due to inequivalent U sites. Starting from the DMFT-derived band structure of the normal state, Ref. (Chen et al., 2021) considered an effective, strong intra-unit cell FM exchange interaction and additional weaker, inter-unit cell AFM exchange couplings as supported by DFT calculations. The resulting spin spectrum in the normal state was found to be in agreement with the AFM fluctuations extracted from INS experiments (Duan et al., 2020; Duan et al., 2021). Following a strong-coupling approach along the lines of an effective, generalized $t$-$J$ model, it was shown that the intra-unit cell FM "dimers" promote a spin-triplet pairing state with nontrivial orbital (layer) DOF. This state also reproduces a resonance in the spin-spectrum, as observed in INS experiments (Duan et al., 2021). In addition, STM measurements also observed anticorrelated modulation between superconducting gap size and Kondo resonance, even within one unit cell (Jiao et al., 2020). Ishizuka and Yanase proposed that applied pressure enhances AFM vs FM fluctuations. The resulting pressure-temperature phase diagram includes an odd-parity spin-triplet state at low pressures, an even-parity spin-singlet phase at intermediate pressures, together with possible mixed odd- and even-parity states in between (Ishizuka and Yanase, 2021).

## 2.2 Parity transition in the superconducting state of CeRh$_2$As$_2$

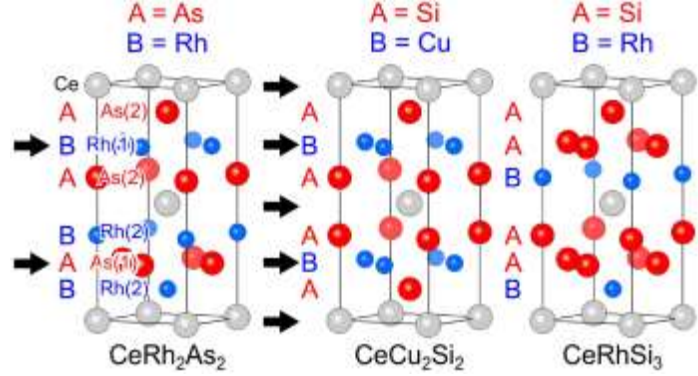

**Figure 6**. Crystal structure of CeRh$_2$As$_2$(left), CeCu$_2$Si$_2$(center), and CeRhSi$_3$(right). To clarify the stacking order, each element is assigned A and B labels. The arrows indicate the cross-section planes of the inversion center. After (Kibune et al., 2021).

The majority of SCs, including HFs, preserve inversion symmetry, which, together with the Pauli exclusion principle, ensures that the Cooper pairs with a singlet (triplet) spin matrix structure must have a corresponding even (odd)-parity wavefunction. Consequently, the pairing states in centrosymmetric SCs can be classified as either even-parity, spin-singlet or odd-parity, spin-triplet. For example, CeCu$_2$Si$_2$ crystallizes in a tetragonal ThCr$_2$Si$_2$-type (*I4/mmm*) structure with inversion symmetry, see Fig. 6 (Kibune et al., 2021), and the Cooper pair instability is believed to occur in the even-parity spin-singlet channel. However, if the crystal structure lacks a global inversion center, an electronic antisymmetric spin-orbit coupling (SOC) is allowed, which in turn lifts the spin degeneracy, inducing Rashba SOC interaction (Gor'kov and Rashba, 2001). Since parity is no longer preserved, a strong antisymmetric SOC can mix spin-singlet and spin-triplet states. Indeed, such parity-mixed superconducting states have been proposed in many non-centrosymmetric compounds such as CePt$_3$Si (Bauer et al., 2004), CeRhSi$_3$ (Kimura et al., 2007), CeCoGe$_3$ (Kawai et al., 2008). These materials can simultaneously exhibit features associated with both *s*-wave and *p*-wave pairing states (Frigeri et al., 2004; Smidman et al., 2017).

Moreover, for some crystal structures, the paired electrons originate from sites which lack a local inversion symmetry, although a global inversion symmetry is preserved. The recently-discovered HF SC CeRh$_2$As$_2$ (with a CaBe$_2$Ge$_2$-type crystal structure) is one example (Khim et al., 2021). A superconducting transition is observed around $T_c = 0.26$ K (Khim et al., 2021; Hafner et al., 2022). While the crystal structure of CeRh$_2$As$_2$ is similar to that of CeCu$_2$Si$_2$, the stacking order differs in the two cases, as shown in Fig. 6. Due to this, in CeRh$_2$As$_2$, the Ce sites lack inversion symmetry. Furthermore, the space-group of the lattice is non-symmorphic. As in a (globally) non-centrosymmetric SC, the absence of local inversion symmetry generates a local electric field (Kinmura et al., 2021). If the spin-orbit interaction is sufficiently strong, the spin degeneracy can be lifted by forming a Rashba spin-texture (Bihlmayer et al., 2015). It has been proposed that the combination of Rashba-type SOC and applied magnetic fields could lead to a change in the parity of the superconducting order parameters in quasi-2D materials, i.e., from even to odd (Yoshida et al., 2012). The observed superconducting states in $CeRh_2As_2$ are seemingly consistent with this simple picture. The large value of specific heat jump at $T_c$ confirms that the superconductivity involves heavy quasiparticles.

In the normal state, $CeRh_2As_2$ shows characteristic Kondo lattice behavior with $T_K \approx 30K$. Low temperature resistivity and specific heat show typical non-Fermi liquid behaviour, indicating that the system is located around a QCP (Khim et al., 2021; Hafner et al., 2022). In addition, the temperature dependence of the NMR relaxation rate demonstrates a singular form for the low-frequency spin dynamics (Kitagawa et al., 2022); the experimental result is compatible with the temperature dependence expected from a KD QCP (Si et al., 2001). All of these observations imply an important role for the AFM fluctuations and Kondo effect in $CeRh_2As_2$, which in this respect is therefore similar to many other HF SC.

For H//c, this compound exhibits very high upper critical fields of ~14 T, more than 20 times larger than the Pauli-limiting field $\mu_0 H_P \sim 1.84\ T_c \sim 0.6$ T. For $H \parallel ab$, the upper critical field is around 1.9 T. The large anisotropic behaviour of $H_{c2}$, which stands in clear contrast to the case of $CeCu_2Si_2$, is however similar to that observed in non-centrosymmetric HF SCs like $CeRhSi_3$. This implies that $CeRh_2As_2$ is a quasi-2D SC with pronounced Rashba SOC. Intriguingly, magnetic and thermodynamic probes unveil an apparent first-order transition around 4 T for $H$//c, as shown in Fig. 7. This anomaly also suggests a change in the symmetry of the superconducting order-parameter with increasing magnetic field. In addition to highlighting the absence of local inversion symmetry, recent experiments also uncovered several other intriguing aspects. 2D AFM fluctuations were observed below a Kondo coherence temperature ~40 K (Kitagawa et al., 2022), with AFM order below $T_c$. The fact that AFM orders occurs within the superconducting phase is quite unusual, with only few examples among thousands of superconductors, such as pressure-induced coexistence of superconducting and AFM order in $CeRhIn_5$ (Park et al., 2006). In $CeRh_2As_2$, this behaviour might be related to the absence of local inversion symmetry at the Ce site, which could induce an effective Zeeman field (Kibune et al., 2021). Slightly above $T_c$, an additional phase transition to a putative non-magnetic, quadrupole density wave was reported (Hafner et al., 2022). Here, Kondo screening and the crystalline-electric-field both play important roles when it comes to determining the low-energy manifold of the Ce $4f$-states. The interplay between superconductivity and quadrupole density waves is still not well understood at this point.

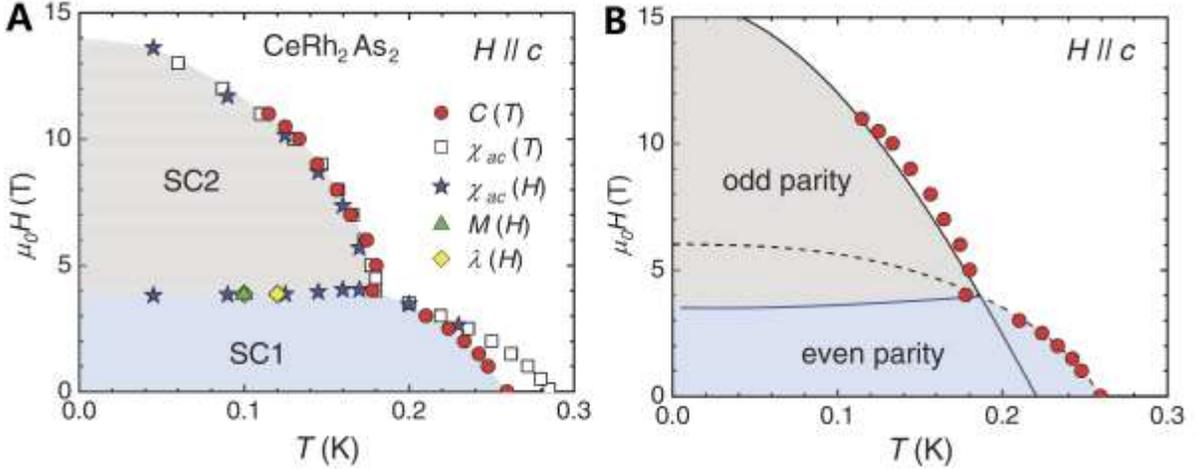

**Figure 7**. Superconducting phase diagrams of CeRh$_2$As$_2$. (A) $H \parallel c$. (B) Fits to the upper critical fields based on an even- to odd-parity transition proposal. After (Khim et al., 2021). [EN: Missing panel B].

Much like for UTe$_2$, the multiple-phase superconductivity in CeRh$_2$As$_2$ (Khim et al., 2021) does not yet allow for a robust consensus on the nature of the Cooper pairing. Nevertheless, we mention those aspects which are unique to CeRh$_2$As$_2$ and discuss existing work in this context. The absence of local inversion symmetry provides a natural connection to previous studies of non-centrosymmetric SCs, where symmetry-allowed Rashba SOC terms can lead to a mixing between even-parity spin-singlet and odd-parity spin-triplet states (Bauer and Sigrist, 2012), as already mentioned.

The two inequivalent effective Ce layers in CeRh$_2$As$_2$ (Fig. 6) can be considered as two non-centrosymmetric superconducting subsystems. Consequently, these layers provide effective orbital DOFs which allow for the construction of matrix pairing states, by analogy to the $s\tau_3$ state discussed in the introduction (Nica et al., 2017; Nica and Si, 2021). Indeed, Ref. (Khim et al., 2021) took this approach. The space of the two inequivalent layers can be denoted by $\tau_0$ and the Pauli matrices $\tau_i$ ($i=1,2,3$). The putative dominant intra-layer, spin-singlet pairing states are classified as even-parity for $\tau_0$ and odd-parity for $\tau_3$ (or equivalently $\tau_z$ in Ref. (Khim et al., 2021)) layer dependencies, with equal- and opposite-sign pairing functions in the two layers, respectively. Due to the Rashba SOC, these states amount respectively to pseudo-spin singlet and helical pseudo-spin triplet states. By virtue of the Zeeman coupling and Cooper pair pseudo-spin orientations, the even-parity state is Pauli limited for fields along the $c$-axis, while the odd-parity state is not. The initial proposal was expanded upon in Ref. (Cavanagh et al., 2022), where the role of the non-symmorphic space group in the context of the Zeeman coupling in the pseudo-spin basis was addressed, together with its consequences for the stability of the even- vs odd-parity states. At the GL level Ref. Schertenleib et al. 2021, considers the Ce layers in CeRh$_2$As$_2$ as two weakly-coupled, non-centrosymmetric superconducting subsystems. For low fields, a predominantly even-parity, spin-singlet pairing state, admixed with a suppressed odd-parity, spin-triplet pairing state with alternating $\pm\pi$ phase along the $c$-axis, is the most stable candidate. For higher fields, Ref. (Schertenleib et al., 2021) finds instead that an odd-parity, spin-triplet state together with a weaker, even-parity, spin-singlet component is favoured instead. These conclusions are in agreement with the experimental results along the $c$-axis.

While the studies mentioned up to this point assumed dominant intra-plane SOC, Ref. (Möckli and Ramires, 2022) considered the effects of stronger inter-layer coupling. In addition to the even-odd parity transition for weak interlayer coupling, the authors also find a distinct transition between two odd-parity spin-triplet states in the opposite limit (Möckli, 2022). Finally, we note that $CeRh_2As_2$ has also been considered in the context of crystalline topological superconductivity protected by non-symmorphic symmetry in Ref. (Nogaki et al., 2021). Among others, the authors advance the notion that odd-parity pairing states, as required by topological superconductivity, can be realized even in the spin-singlet case, due to the alternating sign of pairing with layer, as proposed for the high-field phase of $CeRh_2As_2$.

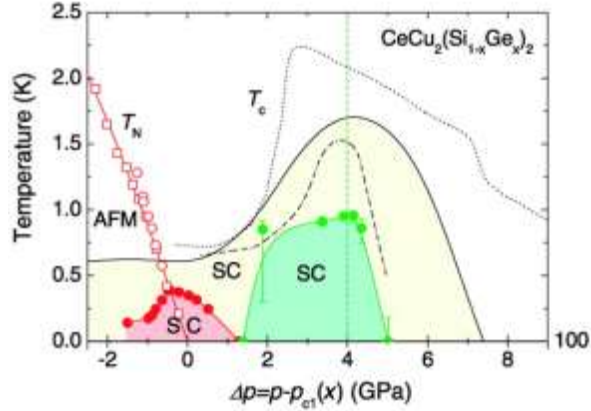

**Figure 8.** Temperature-pressure phase diagram of $CeCu_2(Si_{1-x}Ge_x)_2$, showing two superconducting phases under pressure (Yuan et al., 2003).

### 2.3 Spin and Charge instabilities in $CeCu_2Si_2$

We have emphasized that superconductivity often develops at the border of charge/spin instabilities. In the vicinity of an AFM order, spin-singlet states are generally favored, ensuring that the angular momentum of the Cooper pairs is even. Therefore, a $d$-wave pairing state is normally expected in this case, as also proposed for cuprate and organic SCs (Stewart 2017). In $CeCu_2Si_2$, the $d$-wave scenario is supported by the absence of a coherence peak in the spin lattice relaxation rate $1/T_1$ from NQR measurements (Ishida et al., 1999). INS suggests $d_{x^2-y^2}-$ symmetry (Eremin et al., 2008; Stockert et al., 2011), while the angular-dependent upper critical field measurements indicate $d_{xy}$ symmetry (Vieyra et al., 2011, Yuan et al., 2003). Overall, these studies suggest $d$-wave symmetry which would typically imply gapless superconductivity. Surprisingly, recent low-temperature specific heat (Kittaka et al., 2014) and London penetration depth (Pang et al., 2018; Yamashita et al., 2017) measurements both indicate fully gapped superconductivity, due to the absence of quasiparticle excitations below ~50 mK. To reconcile these experimental results, $s^{\pm}$ and $d+d$ type paring states have been proposed, respectively. A summary of the extensive work on $CeCu_2Si_2$, which stretches back for a few decades, is given in Ref. (Smidman et al., 2018, Smidman et al., 2022).

As discussed in a broader context (Sec. 1.3), including that of the Fe-based SCs, the orbital DOF can play a leading role for Cooper paring. This is also likely the case for HFs in general, which typically exhibit multiple bands of varying orbital content in the vicinity of the Fermi level. The difficulties in resolving the experimental results in $CeCu_2Si_2$ via single-band pairing states naturally suggest that the pool of available candidates must be enlarged. One way of achieving this is to consider the nontrivial effects of additional bands, or equivalently, that of additional local orbital/spin-orbital DOF in the context of pairing states. To illustrate this approach, we review a recently proposed, two-band $d + d$ pairing state, consisting of distinct intra- and inter-band $d$-wave components in 2D (Nica and Si, 2021). This candidate has provided a resolution to the experiments in CeCu$_2$Si$_2$ (Pang et al., 2018; Smidman et al., 2018, Smidman et al., 2022). In contrast to single-band $d$-wave pairing, $d + d$ pairing exhibits a nontrivial matrix structure:

$$\Delta_{d+d} = \begin{pmatrix} d_{x^2-y^2} & d_{xy} \\ d_{xy} & -d_{x^2-y^2} \end{pmatrix}$$

Thus, while the intra- and inter-band components have nodes along the diagonals and axes of the BZ, much like two, distinct simple $d$-waves, their superposition is significantly different. Most notably, due to the matrix structure, the pairing gap is determined by the addition in quadrature of the two components. Since these are distinct $d$-waves, for instance $d_{x^2-y^2}$ and $d_{xy}$, their nodes overlap only at certain high-symmetry points in the BZ. In general, this leads to a fully gapped FS. However, this pairing state also retains crucial aspects of single-band $d$-wave pairing. Thus far, the most attractive feature is the change in sign of the intra-band component under $C_{4z}$ rotations, which accounts for the presence of an in-gap resonance in the spin spectrum, as captured by INS experiments in CeCu$_2$Si$_2$ (Stockert et al., 2011) as well as in the alkaline Fe-selenides (Park et al., 2011; Friemel et al., 2012). The $d + d$ pairing state can therefore 'seamlessly' combine features associated with either single-band $s$- or $d$-waves, considered mutually exclusive in the standard approach. Another important feature of $d + d$ pairing is its relation to microscopic pairing states defined in terms of orbital DOF, which belong to a single irreducible representation of the PG. Two important consequences follow from this statement. First, the two components are not expected to develop independently, as in a multi-gap pairing state (Nagamatsu et al., 2001), implying a single transition to a $d + d$ superconducting state, as for single-band $d$-wave pairing. The $d+d$ pairing state preserves all of the symmetries of the former. Secondly, both $d + d$ and single-band $d$-wave pairing states are classified according to the same irreducible representation by hypothesis. Consequently, $d + d$ pairing can coexist with a single-band $d$-wave pairing, again without breaking any additional symmetries. This includes regimes where $d + d$ pairing is not the strongest superconducting channel; its presence being favored instead by the additional opening of a gap in the BdG spectrum.

While pairing with nontrivial matrix structure provides a new and general way of bypassing the mutually-exclusive nature of single-band $s$- and $d$-wave states, its associated pairing mechanisms and regimes of stability are more conveniently considered in an equivalent representation, involving electrons in local orbital/spin-orbital multiplets. These DOF provide a natural basis for constructing pairing states from correlated electrons subject to strong Coulomb repulsion. As discussed in Sec. 1.3, the idea is that pairing states with a nontrivial matrix structure in an orbital/spin-orbital basis, which transform nontrivially under the PG, lead to pairing states with nontrivial structure in the band basis, like for the $d + d$ state. This is due to the interplay between the local multiplet structure and the normal-state dispersion. The latter is determined by effective intra- and inter-orbital hybridization terms which generally do not commute with nontrivial matrix pairing states. This ensures the presence of both intra- and inter-band components in the band basis, although the pairing state preserves those features associated with PG symmetry. This was explicitly shown in the context of two-orbital models for the alkaline Fe-selenides (Sec. 1.3). The equivalence of $d + d$ pairing to a matrix pairing state consisting of an $s$-wave form factor multiplied by a $\tau_3$ Pauli matrix, defined in the basis of $d_{xz/yz}$ degenerate orbitals, was shown in Ref. (Nica et al., 2017). The sign-changing nature of the $d + d$ state is attributed to the $\tau_3$ pairing matrix, which belongs to a nontrivial $B_{1g}$ irreducible representation of the PG. Moreover, this state was also stabilized in a five-orbital model for the alkaline Fe-selenides. A microscopic candidate for $d+d$ pairing in CeCu$_2$Si$_2$ was advanced along similar lines in Ref. (Nica and Si, 2021). By analogy with the alkaline Fe-selenides, this candidate was labeled as $s\Gamma_3$. It consists of an $s$-wave form factor multiplied by a matrix defined in terms of $\Gamma_7$ correlated f-electron and $\Gamma_6$ conduction electron doublets.

Charge instabilities can also enhance effective pairing interactions (Monthoux et al., 2007). $CeCu_2Si_2$ and its isostructural compound $CeCu_2Ge_2$ both exhibit a superconducting dome around the border of an AFM QCP, see Fig. 8. However, by tuning the samples away from the QCP using pressure, superconductivity is first (completely) suppressed, but subsequently re-emerges to form a secondary dome in the pressure-temperature phase diagram. Several models have been proposed to explain the second dome superconductivity in $CeCu_2Si_2$ and $CeCu_2Ge_2$ (Weng et al., 2016). The point of these proposals is that pressure induces a change of a Ce valence transition, which is more likely associated with an effective charge instability, instead of spin. A number of experimental results corroborate this picture. These include a reduction in the electronic scattering and a maximum Sommerfeld coefficient around the center of the second superconducting dome, as detected from transport measurements (Holmes et al., 2004; Yuan et al., 2003), and a continuous change in the Ce valence for $CeCu_2Si_2$ under increasing pressure, as observed by x-ray absorption spectroscopy (Rueff et al., 2011). Recent research finds additional examples which appear to deviate from the central theme of spin fluctuation-mediated Cooper pairing. For instance, superconductivity and antiferromagnetism can coexist over a wide pressure range in $CeAu_2Si_2$ (Ren et al., 2014), with both superconducting and magnetic transition temperatures showing a parallel increase with pressure. The pressure-temperature phase diagram also resembles those of $CeCu_2Si_2$ and $CeCu_2Ge_2$.

## 3    Summary and Outlook

While single-phase, spin-singlet superconductivity emerging near an AFM QCP still probably provides the most robust understanding of unconventional superconductivity, it is by no means unique, as illustrated in previous sections. For instance, in $UTe_2$, one could argue that the traditional competition between antiferromagnetism and Kondo screening must also include the effects of FM fluctuations. This complicated interplay likely underlies the complex superconducting phase diagram, which includes up to five distinct regimes, as accessed by applying magnetic field and pressure. Furthermore, additional multi-orbital or equivalent microscopic DOF, which emerge from the interplay of lattice structure and symmetry of the $f$-electron sites, can significantly enrich the emerging superconducting phase diagrams. In $CeRh_2As_2$, magnetic fields can in principle induce a change in the parity of the superconducting order parameter from odd to even, in a scenario made plausible by the lack of local inversion symmetry due to stacking order. The two-domed phase diagram of $CeCu_2Si_2$ suggests that spin and charge instabilities can in turn play leading roles in driving multiple superconducting phases in the same compound, as accessed via distinct tuning parameters. In general, we have emphasized how quantum criticality and the associated strange-metal physics promote unconventional superconductivity in heavy-fermion metals. Nevertheless, many aspects of the paring symmetries for these HF superconductors remain to be clarified and further studies are highly desired.

The common multi-orbital/band nature of HFs can play an important role in determining the paring symmetry. Consider such multi-band pairing states such as $s\tau_3$ or, equivalently, $d+d$. In contrast to the intra-band component, the inter-band component in general pairs electrons away from the FS, with a characteristic energy difference determined by the Kondo energy scale. As alluded to in the introduction, and analogous to the case for the alkaline Fe-selenide superconductors discussed earlier, the inter-band pairing is competitive if the pairing interaction is on the scale of the Kondo energy. This is natural, because the heavy fermion superconductors we have covered are either at, or not too far away from a quantum critical point, where the RKKY interaction is expected to be comparable to the Kondo energy scale. This can be seen more explicitly from the experimental measurements of the magnetic fluctuation spectrum. Inelastic neutron scattering experiments, in $CeCu_2Si_2$, for example have identified an energy width in the magnetic excitation spectrum that is on the order of the Kondo energy scale (Stockert et al., 2011). A related issue is the anisotropy of the RKKY interactions, which is typically pronounced in HF compounds with strong local SOC. This is important in several ways. The crystal field splits the total-$J$ multiplets of the quasi-localized $f$-electrons due to strong SOC. The lowest-energy, local manifold determines the nature of the spin-orbital DOFs which can be used to construct unconventional microscopic pairing states with matrix structure. For instance, in $CeCu_2Si_2$, the $f$-electrons are in a $\Gamma_7$ doublet state. The only allowed, nontrivial, even-parity, matrix pairing state involves additional $\Gamma_6$ doublets (Nica and Si, 2021). Even when SOC is not dominant, it can lead to a coupling between the orbital and spin DOF of the pairing state leading to nontrivial structure for given spin-singlet or -triplet configurations, as proposed for $UTe_2$ (Chen et al., 2021). Furthermore, an Ising anisotropy of the RKKY interactions due to SOC, can also make spin-triplet pairing states competitive, even in the nominally AFM case (Hu et al., 2021b). While on the topic of the interplay between orbital and spin DOFs, as promoting pairing states with novel properties, we briefly mention that it can also be central to the important family of the ruthenates. We refer the readers to, for example, a number of ongoing theoretical studies (Ramires, 2021; Lindquist et al., 2022; Puetter and Kee, 2012).

Symmetry-protected topological pairing states have also received a great deal of attention in recent years. As noted previously, spin-triplet pairing in $^3$He has provided important precedents in this context as well, particularly in the case of the gapped and topologically nontrivial B phase with Majorana edge modes, see for instance Refs. (Murakawa et al., 2011; Bunkov et al., 2020) and work cited therein, but also as an early realization of Weyl fermions in the A phase (Volovik 2003). There are some similarities between the phase diagrams of $^3$He and the HF superconductor $UPt_3$ (Vollhardt and Wölfle, 1990; Joynt and Taillefer, 2002). $UTe_2$ has also emerged as a promising HF candidate to spin-triplet superconductivity. Signatures of TRS breaking (Hayes et al., 2021) as well as reports of edge modes (Jiao et al., 2020) have also singled out this compound as a potential realization of topologically nontrivial (chiral) superconductivity (Aoki et al., 2022). $CeRh_2As_2$ has also been advanced as a candidate for crystalline topological superconductivity protected by the non-symmorphic space group (Kosuke et al., 2021). The required odd-parity pairing is due to a matrix pairing structure (in an effective orbital DOF) instead of the more traditional spin-triplet candidates. In the context of the much-sought topological superconductivity, the larger space of pairing candidates which goes hand-in-hand with multiple superconducting phases in a number of HFs is very promising, for predicting and ultimately realizing these exotic states. As illustrated by these recent developments, pairing with matrix structure is inherently richer than the effectively single-band candidates (Nica et al., 2017; Nica and Si, 2021).

Although HFs have been the focus of this work, it is important to note that multiple superconducting phases have also been observed in $YBa_2Cu_3O_y$ (Grissonnanche et al., 2014), $LaFeAsO_{1-x}(H/F)_x$ (Iimura et al., 2012), $BaTi_2(Sb_{1-x}Bi_x)_2O$ (Zhai et al., 2013), and very recently, in the kagome superconductors $CsV_3Sb_5$ (Nguyen et al., 2022) and $RbV_3Sb_5$ (Guguchia et al., 2022). As illustrated here, recent advances in experimental techniques have provided a much-improved characterization of sample properties in transport, thermodynamic, and various spectroscopies. In such cases as $UTe_2$ and $CeRh_2As_2$, these advances have significantly accelerated the discovery of multiple superconducting phases. This progress has also led to dramatic reversals in so-called "standard" cases. Indeed, owing to the development of precise measurements below 100 mK, a completely new understanding of $CeCu_2Si_2$, the very first unconventional superconductor is now emerging, with possible implications for other classes of well-studied superconductors. The great diversity of unconventional superconductors provides unprecedented opportunities for realizing novel pairing states.

**Conflict of Interest**

The authors declare that the research was conducted in the absence of any commercial or financial relationships that could be construed as a potential conflict of interest.

**Author Contributions**

All authors contributed to the writing of the manuscript.

**Funding**

E.M.N. was supported by the NSF Grant No. DMR-1904716 and an ASU startup grant. The work at Washington University was supported by McDonnell International Scholars Academy. L.J. acknowledges financial support from National High Magnetic Field Laboratory, which is funded by the National Science Foundation (DMR-1644779) and the State of Florida. The work at Rice University has primarily been supported by the DOE BES Award #DE-SC0018197; additional support has been provided by the Robert A. Welch Foundation Grant No. C-1411.

**Acknowledgments**

We thank Lei Chen, Pengcheng Dai, Chunruo Duan, Onur Erten, Haoyu Hu, Silke Paschen, Frank Steglich, Rong Yu, Vidya Madhavan, and Steffen Wirth for useful discussions.

## References

Adenwalla, S., Lin, S. W., Ran, Q. Z., Zhao, Z., Ketterson, J. B., Sauls, J. A. et al., (1990). Phase diagram of UPt$_3$ from ultrasonic velocity measurements. Phys. Rev. Lett. 65, 2298.

Aoki, D. Huxley A., Ressouche, E., Braithwaite, D., Flouquet, J., Brison, J. -P., et al. (2001). Coexistence of superconductivity and ferromagnetism in URhGe. Nature 413, 613.

Aoki, D., Nakamura, A., Honda, F., Li, D. X., Homma, Y., Shimizu, Y. et al. (2019). Unconventional superconductivity in heavy fermion UTe$_2$. Journal of the Physical Society of Japan 88, 043702. https://doi.org/10.7566/JPSJ.88.043702.

Aoki, D., Honda, F., Knebel, G., Braithwaite, D., Nakamura, A., Li, D. X., et al. (2020). Multiple superconducting phases and unusual enhancement of the upper critical field in UTe$_2$. J. Phys. Soc. Japan 89, 053705.

Aoki, D., Brison, J. -P., Flouquet, J., Ishida, K., Knebel, G., Tokunaga, Y., and Yanase, Y. (2022) Unconventional Superconductivity in UTe$_2$, J. Phys.: Condens. Matter. https://iopscience.iop.org/article/10.1088/1361-648X/ac5863

Arndt, J., Stockert, O., Schmalzl, K., Faulhaber, E., Jeevan, H. S., Geibel, C., Schmidt, W., Loewenhaupt, M., and Steglich, F. (2011). Spin Fluctuations in Normal State CeCu$_2$Si$_2$ on Approaching the Quantum Critical Point. Phys. Rev. Lett. 106, 246401.

Bae, S. Kim, H. S., Eo, Y. S., Ran, S., Liu, I, Fuhrman. W. T., et al. (2021). Anomalous normal uid response in a chiral superconductor UTe$2$. Nat. Commun. 12, 2644. https://doi.org/10.1038/s41467-021-22906-6.

Bardeen, J., Cooper, L. N., and Schrieffer J. R. (1957). Microscopic theory of superconductivity, Phys. Rev. 106 162.

Bauer, E., Hilscher, G., Michor, H., Paul, C., Scheidt, E. W., Gribanov, A., et al., (2004). Phys. Rev. Lett. 92 027003.

Bauer E., and Sigrist. M. (2012). Non-Centrosymmetric Superconductors: Introduction and Overview. Lecture notes in physics. Springer-Verlag Berlin Heidelberg.

Bianchi. A., Movshovich, R., Vekhter, I., Pagliuso, P. G., and Sarrao, J. L. (2003). Avoided Antiferromagnetic Order and Quantum Critical Point in CeCoIn$_5$. Phys. Rev. Lett. 91, 257001.

Bihlmayer, G., Rader, O., and Winkler, R., (2015). Focus on the Rashba effect. New J. Phys. 17, 050202.

Braithwaite, D., Vališka, M., Knebel, G., Lapertot, G., Brison, J. -P., Pourret, A., et al. (2019). Multiple superconducting phases in a nearly ferromagnetic system. Commun. Phys. 2, 147. URL https://doi.org/10.1038/s42005-019-0248-z.

Bucher, E., Maita, J. P., Hull, G. W., Fulton, R. C., and Cooper, A. S. (1975). Electronic properties of beryllides of the rare earth and some actinides. Phys. Rev. B 11, 440. DOI:10.1103/PhysRevB.11.440.

Bunkov Y. M., and Gazizulin, R. R. (2020). Direct observation of the specific heat of Majorana quasiparticles in superfluid $^3$He B phase. Scientific Reports 10, 20120.

Cai, A., Pixley, J. H., Ingersent, K., and Si, Q. (2020). Critical local moment fluctuations and enhanced pairing correlations in a cluster Anderson model. Phys. Rev. B 101, 014452.

Cavanagh, D. C., Shishidou, T., Weinert, M., Brydon, P. M. R., and Agterberg, D. F. (2022). Nonsymmorphic symmetry and field-driven odd-parity pairing in CeRh$_2$As$_2$. Phys. Rev. B 105, L020505.

Chen, L., Hu, H., Lane, C., Nica, E. M., Zhu, J. X., and Si, Q. (2021). Multiorbital spin-triplet pairing and spin resonance in the heavy-fermion superconductor UTe$_2$. arXiv:2112.14750.

Coleman, P., Pépin, C., Si, Q., and Ramazashvili, R. (2001). How do Fermi liquids get heavy and die? J. Phys.: Condens. Matter 13, R723-R738.

Coleman, P., Schofield, A. J. (2005). Quantum criticality.  Nature 433, 226-229.

Custers, J., Gegenwart, P., Wilhelm, H., Neumaier, K., Tokiwa, Y., Trovarelli, O., et al. (2003), The break-up of heavy electrons at a quantum critical point. Nature 424, 524.

Custers, J., Lorenzer, K., Müller, M., Prokofiev, A., Sidorenko, A., Winkler, H., Strydom, A. M., Shimura, Y., Sakakibara, T., Yu, R., Si, Q., and  Paschen, S. (2012). Destruction of the Kondo effect in the cubic heavy-fermion compound Ce$_3$Pd$_{20}$Si$_6$. Nat. Mater. 11, 189-194.

Dai, P. C. (2015) Antiferromagnetic order and spin dynamics in iron-based superconductors. Rev. Mod. Phys. 87, 855–896.

Doniach, S. (1977). The Kondo lattice and weak antiferromagnetism. Physica B + C 91, 231.

Duan, C., Sasmal, K., Maple, M. B., Podlesnyak, A., Zhu, J.-X., Si, Q., and Dai, P. (2020). Incommensurate Spin Fluctuations in the Spin-Triplet Superconductor Candidate UTe$_2$. Phys. Rev. Lett. 125, 237003.

Duan, C. R., Baumbach, R. E., Podlesnyak, A., Deng, Y. H., Moir, C., Breindel, A. J., Maple, B., Nica, E. M., Si, Q., and Dai, P. C. (2021). Resonance from antiferromagnetic spin fluctuations for superconductivity in UTe$_2$. Nature 600, 636–640.

Eremin, I., Zwicknagl, G., Thalmeier, P., and Fulde, P., 2008. Feedback Spin Resonance in Superconducting CeCu$_2$Si$_2$ and CeCoIn$_5$. Phys. Rev. Lett. 101 187001

Ernst, S., Kirchner, S., Krellner, C., Geibel, C., Zwicknagl, G., Steglich, F., and Wirth, S. (2011). Emerging local Kondo screening and spatial coherence in the heavy-fermion metal YbRh$_2$Si$_2$. Nature 474, 362.

Friedemann, S., Oeschler, N., Wirth, S., Krellner, C., Geibel, C., Steglich, F., Paschen, S., Kirchner, S., and Si, Q. (2010). Fermi-surface collapse and dynamical scaling near a quantum-critical point," Proc. Natl. Acad. Sci. USA 107, 14547–14551.
https://www.pnas.org/doi/pdf/10.1073/pnas.1009202107.

Friemel, G., Park, J. T., Maier, T. A., Tsurkan, V., Li, Y., Deisenhofer, J. et al. (2012). Reciprocal-space structure and dispersion of the magnetic resonant mode in the superconducting phase of $Rb_xFe_{2-y}Se_2$ single crystals. Phys. Rev. B 85, 140511.

Frigeri, P. A., Agterberg, D. F., and Sigrist, M. (2004). Spin susceptibility in superconductors without inversion symmetry. New J. Phys. 6 115.

Gegenwart, P., Custers, J., Geibel, C., Neumaier, K., Tayama, T., Tenya, K., Trovarelli, O., and Steglich, F. (2002). Magnetic-Field Induced Quantum Critical Point in $YbRh_2Si_2$. Phys. Rev. Lett. 89, 056402.

Gegenwart, P., Si, Q., and Steglich, F. (2008). Quantum criticality in heavy-fermion metals, Nat. Phys. 4 186.

Gerber, S., Bartkowiak, M., Gavilano, J. L., Ressouche, E., Egetenmeyer, N., Niedermayer, C., et al. (2014). Switching of magnetic domains reveals spatially inhomogeneous superconductivity, Nat. Phys. 10, 126.

Gor'kov, L. P., and Rashba, E. I. (2001). Superconducting 2D System with Lifted Spin Degeneracy: Mixed Singlet-Triplet State. Phys. Rev. Lett. 87 037004.

Grissonnanche, G., Cyr-Choinière, O., Laliberté, F., René de Cotret, S., Juneau-Fecteau, A., Dufour-Beauséjour, S. et al. (2014). Direct measurement of the upper critical field in cuprate superconductors. Nat. Commun. 5 3280.

Grosche, F. M., Walker, I. R., Julian, S. R., Mathur, N. D., Freye, D. M., Steiner, M. J., and Lonzarich, G. G., (2001). Superconductivity on the threshold of magnetism in $CePd_2Si_2$ and $CeIn_3$. J. Phys.: Condens. Matter 13 2845–2860

Guguchia, Z., Mielke, C. III, Das, D., Gupta, R., Yin, J.-X., Liu, H. et al. (2022). Tunable nodal kagome superconductivity in charge ordered $RbV_3Sb_5$. arXiv:2202.07713

Hafner, D., Khanenko, P., Eljaouhari E.-O., Kuchler, R., Banda, J., Bannor, N.. et al. (2022). Possible Quadrupole Density Wave in the Superconducting Kondo Lattice CeRh2As2. Phys. Rev. X 12, 011023

Hayes, I. M. Wei, D. S., Metz, T., Zhang, J., Eo, Y. S., Ran, S., et al. (2021). Multicomponent superconducting order parameter in $UTe_2$. Science 373, 797-801. DOI: 10.1126/science.abb0272

Hertz, J. A. (1976). Quantum critical phenomena. Phys. Rev. B 14, 1165-1184.

Holmes, A. T., Jaccard, D., and Miyake, K. 2004. Signatures of valence fluctuations in $CeCu_2Si_2$ under high pressure. Phys. Rev. B 69 024508

Hu, H., and Si, Q. (2020). Talk given at the January 2020 fundamentals of quantum materials workshop on superconductivity of $UTe_2$, University of Maryland.

Hu, H., Cai, A., Chen, L., Deng, L., Pixley, J. H., Ingersent, K., and Si, Q. (2021a). Unconventional Superconductivity from Fermi Surface Fluctuations in Strongly Correlated Metals. arXiv:2109.13224.

Hu, H., Cai, A., Chen, L., and Si, Q. (2021b). Spin-singlet and spin-triplet pairing correlations in antiferromagnetically coupled kondo systems. arXiv:2109.12794.

Huy, N. T., Gasparini, A., Nijs, D. E. de, Huang, Y., Klaasse, J. C. P., Gortenmulder, T., et al. (2007). Superconductivity on the Border of Weak Itinerant Ferromagnetism in UCoGe. Phys. Rev. Lett. 99, 067006.

Huxley, A., Rodière, P., Paul, D. M., Dijk, N. v., Cubitt, R., and Flouquet, J. (2000). Realignment of the flux-line lattice by a change in the symmetry of superconductivity in $UPt_3$. Nature 406, 160-164.

Huxley, A., Sheikin, I., Ressouche, E., Kernavanois, N., Braithwaite, D., Calemczuk, R., and Flouquet, J. (2001), $UGe_2$: A ferromagnetic spin-triplet superconductor, Phys. Rev. B 63, 144519.

Iimura, S., Matsuishi, S., Sato, H., Hanna, T., Muraba, Y., Kim, S. W. et al. (2012). Two-dome structure in electron-doped iron arsenide superconductors. Nat. Commun. 3 943.

Ishida, K., Kawasaki, Y., Tabuchi, K., Kashima, K., Kitaoka, Y., Asayama, K., et al. (1999). Evolution from Magnetism to Unconventional Superconductivity in a Series of $Ce_xCu_2Si_2$ Compounds Probed by Cu NQR. Phys. Rev. Lett. 82 5353.

Ishizuka, J., Sumita, S., Daido, A., and Yanase, Y. (2019). Insulator-metal transition and topological superconductivity in $UTe_2$ from a first-principles calculation," Phys. Rev. Lett. 123, 217001.

Jiao, L., Howard, S., Ran, S., Wang, Z. Y., Rodriguez, J. O., Sigrist, M., et al. (2020). Chiral superconductivity in heavy-fermion metal $UTe_2$. Nature 579, 523-527. URL https://doi.org/10.1038/s41586-020-2122-2.

Joynt, R., and Taillefer L. (2002). The superconducting phases of $UPt_3$. Rev. Mod. Phys. 74, 235–294.

Kang, K., Fernandes, R. M., Abrahams, E., and Wölfle, P. (2018). Superconductivity at an antiferromagnetic quantum critical point: Role of energy fluctuations," Phys. Rev. B 98, 214515.

Kawai, T., Muranaka, H., Measson, M., Shimoda, T., Doi, Y., Matsuda T. D. et al., (2008). Magnetic and Superconducting Properties of Ce$TX_3$ ($T$: Transition Metal and $X$: Si and Ge) with Non-centrosymmetric Crystal Structure.

Kawasaki, S., Yashima, M., Mugino, Y., Mukuda, H., Kitaoka, Y., Shishido, H., and Ōnuki, Y. (2006). Enhancing the Superconducting Transition Temperature of $CeRh_{1−x}Ir_xIn_5$ due to the Strong-Coupling Effects of Antiferromagnetic Spin Fluctuations: An [115]In Nuclear Quadrupole Resonance Study. Phys. Rev. Lett. 96 147001.

Kibune, M., Kitagawa, S., Kinjo, K., Ogata, S., Manago, M., Taniguchi, T., et al., (2021). Observation of antiferromagnetic order as odd-parity multipoles inside the superconducting phase in $CeRh_2As_2$. arXiv:2112.07081.

Kim, D. Y.，Lin, S. Y., Weickert, F., Kenzelmann, M., Bauer, E. D., Ronning, F., et al. (2016). Intertwined Orders in Heavy-Fermion Superconductor $CeCoIn_5$. Phys. Rev. X, 6, 041059.

Kimura, N., Ito, K., Aoki, H., Uji, S., and Terashima, T. (2007). Extremely High Upper Critical Magnetic Field of the Noncentrosymmetric Heavy Fermion Superconductor CeRhSi$_3$. Phys. Rev. Lett. 98 197001.

Kimura, S., Sichelschmidt, J., and Khim, S. Y. (2021). Optical study on electronic structure of the locally non-centrosymmetric CeRh$_2$As$_2$. arXiv:2109.00758.

Kirchner, S., Paschen, S., Chen, Q., Wirth, S., Feng, D., Thompson, J. D., and Si, Q. (2020). Colloquium: Heavy-electron quantum criticality and single-particle spectroscopy. Rev. Mod. Phys. 92, 011002.

Kitagawa, S., Kibune, M., Kinjo, K., Manago, M., Taniguchi, T., Ishida, K., Brando, M., Hassinger, E., Geibel, C., and Khim, S. (2022). Two-dimensional XY-type magnetic properties of locally noncentrosymmetric superconductor CeRh$_2$As$_2$. J. Phys. Soc. Japan 91, 043702.

Kittaka, S., Aoki, Y., Shimura, Y., Sakakibara, T., Seiro, S., Geibel, C., Steglich, F., Ikeda, H., and Machida, K., (2014). Multiband superconductivity with unexcepted deficiency of nodal quasiparticles in Cecu$_2$Si$_2$. Phys. Rev. Lett. 112 067002.

Khim, S., Landaeta, J. F., Banda, J., Bannor, N., Brando, M., Brydon, P. M. R., et al. (2021). Field-induced transition within the superconducting state of CeRh$_2$As$_2$. Science 373, 1012–1016. https://www.science.org/doi/pdf/10.1126/science.abe7518

Knafo, W. Nardone, M., Vališka, M., Zitouni, A., Lapertot, G., Aoki, D., Knebel G., and Braithwaite D. (2021). Comparison of two superconducting phases induced by a magnetic field in UTe$_2$. Commun. Phys. 4, 40. URL https://doi.org/10.1038/s42005-021-00545-z.

Knebel, G., Aoki, D., Brison, NJ. -P., and Flouquet, J. (2008). The quantum critical point in cerhin5: A resistivity study. Journal of the Physical Society of Japan 77, 114704. https://doi.org/10.1143/JPSJ.77.114704.

Knebel, G. Knafo, W., Pourret, A., Niu, Q., Vališka, M., Braithwaite D., et al. (2019). Field-reentrant superconductivity close to a metamagnetic transition in the heavy-fermion superconductor UTe$_2$. J. Phys. Soc. Jpn. 88, 063707. URL https://doi.org/10.7566/JPSJ.88.063707.

Kreisel, A., Quan, Y. D., and Hirschfeld, P. J. (2022). Spin triplet superconductivity driven by finite-momentum spin fluctuations. Phys. Rev. B 105, 104507.

Lee, D.-H. (2017). Hunting down unconventional superconductors. Science 357, 32-33.

Leggett A. J. (1975). A theoretical description of the new phases of liquid $^3$He. Rev. Mod. Phys. 47, 331–414.

Li, Y., Wang, Q. Q., Xu, Y. J., Xie, W. H., and Yang, Y. F., (2019). Nearly degenerate p$_x$+i$p_y$ and dx2-y2 pairing symmetry in the heavy fermion superconductor YbRh$_2$Si$_2$. Phys. Rev. B 100, 085132.

Lin, W.-C., Campbell, D. J., Ran, S., Liu, I., Kim, H. S., Nevidomskkyy, A. H., et al. (2020). Tuning magnetic confinement of spin-triplet superconductivity. npj Quantum Materials 5, 68. URL https://doi.org/10.1038/s41535-020-00270-w.

Lindquist, A. W. and Clepkens, J. and Kee, H.-Y., (2022). Evolution of interorbital superconductor to intraorbital spin-density wave in layered ruthenates. Phys. Rev. Research 4, 023109.

Löhneysen, H. v., Pietrus, T., Portisch, G., Schlager, H. G., Schröder, A., Sieck, M., and Trappmann, T. (1994). Non-fermi-liquid behavior in a heavy-fermion alloy at a magnetic instability. Phys. Rev. Lett. 72, 3262–3265.

Löhneysen, H. v., Rosch, A., Vojta, M., and Wölfle, P. (2007). Fermi-liquid instabilities at magnetic quantum critical points. Rev. Mod. Phys. 79, 1015-1075.

Martelli, V., Cai, A., Nica, E. M., Taupin, M., Prokofiev, A., Liu, C.-C., Lai, H.-H., Yu, R., Ingersent, K., Küchler, R., Strydom, A. M., Geiger, D., Haenel, J., Larrea, J., Si, Q., and Paschen, S. (2019). Sequential localization of a complex electron fluid. Proc. Natl. Acad. Sci. U.S.A. 116, 17701.

Mathur, N. D., Grosche, F. M., Julian, S. R., Walker, I. R., Freye, D. M., Haselwimmer, R. K. W., and Lonzarich, G. G. (1998). Magnetically mediated superconductivity in heavy fermion compounds. Nature 394, 39-43.

Matsubayashi, K., Tanaka, T., Sakai, A., Nakatsuji, S., Kubo, Y., Uwatoko, Y. (2012). Pressure-induced heavy fermion superconductivity in the nonmagnetic quadrupolar system $PrTi_2Al_{20}$. Phys. Rev. Lett. 109 187004.

Metz, T. Bae, S. J., Ran, S., Liu, I., Eo, Y. S., Fuhrman, W. T., et al. (2019). Point-node gap structure of the spin-triplet superconductor $UTe_2$. Phys. Rev. B 100, 220504. https://link.aps.org/doi/10.1103/PhysRevB.100.220504.

Miao, L., Liu, S. Z., Xu, Y. S., Kotta, E. C., Kang, C. -J., Ran, S., et al. (2020). Low energy band structure and symmetries of $UTe_2$ from angle-resolved photoemission spectroscopy," Phys. Rev. Lett. 124, 076401.

Millis, A. J. (1993). Effect of a nonzero temperature on quantum critical points in itinerant fermion systems. Phys. Rev. B 48, 7183-7196.

Miyake, A., Shimizu, Y., Sato, Y. J., Li, D. X., Nakamura, A., Homma, Y., et al. (2019). Metamagnetic transition in heavy fermion superconductor $UTe_2$. J. Phys. Soc. Jpn. 88, 063706. URL https://doi.org/10.7566/JPSJ.88.063706.

Miyake, K., Schmitt-Rink, S., and Varma, C. M. (1986). Spin-fluctuation-mediated even-parity pairing in heavy-fermion superconductors. Phys. Rev. B 34, 6554.

Monthoux, P., and Lonzarich, G. G. (2001). Magnetically mediated superconductivity in quasi-two and three dimensions. Phys. Rev. B 63, 054529.

Monthoux, P., Pines, D., and Lonzarich, G. G. (2007). Superconductivity without phonons, Nature 450, 1177.

Movshovich, R., Graf, T., Mandrus, D., Thompson, J. D., Smith, J. L., and Fisk, Z. (1996). Superconductivity in heavy-fermion $CeRh_2Si_2$. Phys. Rev. B 53, 8241

Murakawa, S., Wada, Y., Tamura, Y., Wasai, M., Saitoh, M., Aoki, Y., et al. (2011). Surface Majorana cone of the superfluid $^3$He B phase. Journal of the Physical Society of Japan 80, 013602. https://doi.org/10.1143/JPSJ.80.013602.

Möckli, D., and Ramires, A. (2021). Two scenarios for superconductivity in CeRh$_2$As$_2$. Phys. Rev. Research 3, 023204.

Möckli, D. (2022). Unconventional singlet-triplet superconductivity. J. Phys.: Conf. Ser. 2164, 12009.

Nagamatsu, J., Nakagawa, N., Muranaka, T., Zenitani, Y., and Akimitsu, J. (2001). Superconductivity at 39 K in magnesium diboride. Nature 410, 63.

Nagato, K., Yamamoto, M., and Nagai, K. (1998). Rough surface effects on the p-wave Fermi superfluids. J. Low Temp. Phys. 110, 1135

Nakamine, G., Kitagawa, S., Ishida, K., Tokunaga. Y., Sakai, H., Kambe, S., et al. (2019). Superconducting properties of heavy fermion UTe$_2$ revealed by $^{125}$Te-nuclear magnetic resonance. J. Phys. Soc. Jpn. 88, 113703. URL https://doi.org/10.7566/JPSJ.88.113703.

Nakamine, G., Kinjo, K., Kitagawa, S., Ishida, K., Tokunaga, Y., Sakai, H., (2021) Anisotropic response of spin susceptibility in the superconducting state of UTe$_2$ probed with $^{125}$Te-NMR measurement. Phys. Rev. B 103, L100503. URL https://link.aps.org/doi/10.1103/PhysRevB.103.L100503.

Nevidomskyy, A. H. (2020). Stability of a nonunitary triplet pairing on the border of magnetism in UTe$_2$. arXiv:2001.02699.

Nguyen, D. H., Sidorenko, A., Taupin, M., Knebel, G., Lapertot, G., Schuberth, E., and Paschen, S. (2021). Superconductivity in an extreme strange metal. Nat. Commun. 12, 4341.

Nguyen, T., and Li, M. D. (2022). Electronic properties of correlated kagomé metals AV$_3$Sb$_5$ (A = K, Rb, and Cs): A perspective. J. Appl. Phys. 131, 060901.

Nica, E. M., Yu, R., and Si, Q. (2017). Orbital-selective pairing and superconductivity in iron selenides. npj Quantum Materials 2, 24.

Nica, E. M., and Si, Q. (2021). Multiorbital singlet pairing and d+d superconductivity. npj Quantum Materials 6, 3.

Nogaki, K., Daido, A., Ishizuka, J., and Yanase, Y. (2021). Topological crystalline superconductivity in locally noncentrosymmetric CeRh$_2$As$_2$. Phys. Rev. Research 3, L032071.

Ott, H. R., Rudigier, H., Fisk, Z., and Smith, J. L. (1985). Phase transition in the superconducting state of U$_{1-x}$Th$_x$Be$_{13}$ ($x$ = 0–0.06) Phys. Rev. B 31, R165.

Palstra, T. T. M., Menovsky, A. A., Berg, J. v., Dirkmaat, A. J., Kes, P. H., Nieuwenhuys, G. J., and Mydosh, J. A. (1985). Superconducting and magnetic transitions in the heavy-fermion system URu$_2$Si$_2$. Phys. Rev. Lett. 55, 2727.

Pang, G. M., Smidman, M., Zhang, J. L., Jiao, L., Weng, Z. F., Nica, E. M., et al. (2018). Fully gapped d-wave superconductivity in CeCu$_2$Si$_2$. Proc. Nat. Acad. Sci. 115, 5343.

Park, T., Ronning, F., Yuan, H. Q., Salamon, M. B., Movshovich, R., Sarrao, J. L., and Thompson, J. D., (2006). Hidden magnetism and quantum criticality in the heavy fermion superconductor CeRhIn$_5$. Nature 440, 65–68.

Park, J. T., Friemel, G., Li, Y., Kim, J. -H. Tsurkan, V. Deisenhofer, J. et al. (2011). Magnetic Resonant Mode in the Low-Energy Spin-Excitation Spectrum of Superconducting Rb$_2$Fe$_4$Se$_5$ Single Crystals. Phys. Rev. Lett. 107, 177005.

Paschen, S., Lühmann, T., Wirth, S., Gegenwart, P., Trovarelli, O., Geibel, C., Steglich, F., Coleman, P., and Si, Q. (2004). Hall-effect evolution across a heavy-fermion quantum critical point. Nature 432, 881–885.

Paschen S., and Si, Q. (2021). Quantum phases driven by strong correlations. Nature Reviews Physics 3, 9–26.

Pixley, J. H., Deng, L. L., Ingersent, K., and Si, Q. (2015). Pairing correlations near a kondo-destruction quantum critical point. Phys. Rev. B 91, 201109.

Prochaska, L., Li, X., MacFarland, D. C., Andrews, A. M., Bonta, M., Bianco, E. F., Yazdi, S., Schrenk, W., Detz, H., Limbeck, A., Si, Q., Ringe, E., Strasser, G., Kono, J., and Paschen, S. (2020). Singular charge fluctuations at a magnetic quantum critical point. Science 367, 285-288.

Puetter, C. M., and Kee, H.-Y. (2012). Identifying spin-triplet pairing in spin-orbit coupled multi-band superconductors. Europhys. Lett. 98, 27010.

Ramires, A. (2021). Nodal gaps from local interactions in Sr$_2$RuO$_4$. J. Phys.: Conf. Ser. 2164, 012002.

Ran, S., Eckberg, C., Ding, Q. -P., Furukawa, Y., Metz, T., Saha, S. R., et al. (2019a). Nearly ferromagnetic spin-triplet superconductivity. Science 365, 684–687. https://www.science.org/doi/pdf/10.1126/science.aav8645.

Ran, S., Liu, I., Eo, Y. S., Campbell, D. J., Neves, P. M., Fuhrman, W. T., et al. (2019b). Extreme magnetic field-boosted superconductivity. Nat. Phys. 15, 1250-1254. https://doi.org/10.1038/s41567-019-0670-x.

Ran, S., Kim, H. S., Liu, I., Saha, S. R., Hayes, I., Metz, T., et al. (2020). Enhancement and reentrance of spin triplet superconductivity in UTe2 under pressure. Phys. Rev. B 101, 140503. URL https://link.aps.org/doi/10.1103/PhysRevB.101.140503.

Ran, S., Saha, S. R., Liu, I., Graf, D., Paglione, J., and Butch, N. P. (2021). Expansion of the high field-boosted superconductivity in UTe$_2$ under pressure. npj Quantum Materials 6, 75. https://doi.org/10.1038/s41535-021-00376-9

Ren, Z., Pourovskii, L. V., Giriat, G., Lapertot, G., Georges, A. and Jaccard, D. (2014). Giant Overlap between the Magnetic and Superconducting Phases of CeAu$_2$Si$_2$ under Pressure. Phys. Rev. X 4 031055.

Rueff, J. -P., Raymond, S., Taguchi, M., Sikora, M., Itié, J. -P., Baudelet, F., et al., 2011. Pressure-Induced Valence Crossover in Superconducting $CeCu_2Si_2$. Phys. Rev. Lett. 106 186405.

Sakai, A., Kuga, K., and Nakatsuji, S. (2012). Superconductivity in the ferroquadrupolar state in the quadrupolar Kondo lattice $PrTi_2Al_{20}$, J. Phys. Soc. Jpn. 81, 083702.

Sauls, J.A. (1994). The order parameter for the superconducting phases of $UPt_3$. Advances in Physics 43, 113–141 https://doi.org/10.1080/00018739400101475.

Schertenleib, E. G., Fischer, M. H., and Sigrist, M. (2021). Unusual *H-T* phase diagram of $CeRh_2As_2$: The role of staggered noncentrosymmetricity. Phys. Rev. Research 3, 023179.

Schröder, A., Aeppli, G., Coldea, R., Adams, M., Stockert, O., Löhneysen, H. v., Bucher, E., Ramazashvili, R., and Coleman, P. (2000). Non-fermi-liquid behavior in a heavy-fermion alloy at a magnetic instability. Nature 407, 351–355.

Schuberth, E., Tippmann, M., Steinke, L., Lausberg, S., Steppke, A., Brando, M., et al. (2016). Emergence of superconductivity in the canonical heavy-electron metal $YbRh_2Si_2$. Science 351, 485–488. https://www.science.org/doi/pdf/10.1126/science.aaa9733.

Senthil, T., Vojta, M., and Sachdev, S. (2004). Weak magnetism and non-Fermi liquids near heavy-fermion critical points. Phys. Rev. B 69, 035111.

Si, Q., Rabello, S., Ingersent, K., and Smith, J. L. (2001). Locally critical quantum phase transitions in strongly correlated metals. Nature 413, 804–808.

Si, Q., and Steglich, F. (2010). Heavy fermions and quantum phase transitions. Science 329, 1161-1166. DOI: 10.1126/science.1191195

Si, Q., Pixley, J. H., Nica, E., Yamamoto ,S. J., Goswami ,P., Yu ,R., and Kirchner ,S. (2014). Kondo Destruction and Quantum Criticality in Kondo Lattice Systems, Journal of the Physical Society of Japan, 83, 061005.

Si, Q., Yu, R., and Abrahams, E. (2016). High-temperature superconductivity in iron pnictides and chalcogenides. Nat. Rev. Mater. 1, 16017.

Sigrist, M., and Ueda, K. (1991). Phenomenological theory of unconventional superconductivity. Rev. Mod. Phys. 63, 239.

Shang, T., Baumbach, R. E., Gofryk, K., Ronning, F.,  Weng, Z. F.,  Zhang, J. L., et al. (2014). $CeIrIn_5$: Superconductivity on a magnetic instability. Phys. Rev. B 89, 041101(R).

Shishido, H., Settai, R., Harima, H., and Ōnuki, Y. (2005). A drastic change of the fermi surface at a critical pressure in $CeRhIn_5$: dHvA study under pressure. Journal of the Physical Society of Japan 74, 1103–1106. https://doi.org/10.1143/JPSJ.74.1103.

Shishidou, T., Suh, H. G., Brydon, P. M. R., Weinert, M., and Agterberg, D. F. (2021). Topological band and superconductivity in $UTe_2$. Phys. Rev. B 103, 104504.

Smidman, M., Stockert, O., Nica,  E. M., Liu, Y., Yuan, H. Q., Si, Q., and Steglich, F. (2022). Unconventional and fully-gapped heavy-fermion superconductivity in CeCu2Si2. Preprint.

Smidman, M., Stockert, O., Arndt, J., Pang, G. M., Jiao, L., H. Q. Yuan, et al. (2018). Interplay between unconventional superconductivity and heavy-fermion quantum criticality: $CeCu_2Si_2$ versus $YbRh_2Si_2$. Philos. Mag. 98, 2930.

Smidman, M., Salamon, M. B., Yuan, H. Q., and Agterberg, D. F. (2017). Superconductivity and spin–orbit coupling in non-centrosymmetric materials: a review. Rep. Prog. Phys. 80 036501.

Steglich, F. (2014). Heavy fermions: superconductivity and its relationship to quantum criticality. Philos. Mag. 94, 3259-3280. doi.org/10.1080/14786435.2014.956835

Steglich, F., Aarts, J., Bredl, C. D., Lieke, W., Meschede, D., Franz, W., and Schäfer, H. (1979). Superconductivity in the Presence of Strong Pauli Paramagnetism: $CeCu_2Si_2$. Phys. Rev. Lett. 43, 1892.

Steglich, F., and Wirth, S. (2016). Foundations of heavy-fermion superconductivity: lattice Kondo effect and Mott physics. Rep. Prog. Phys. 79, 084502.

Stewart G. R. (2017). Unconventional superconductivity. Advances in Physics 66, 75–196 https://doi.org/10.1080/00018732.2017.1331615.

Stewart, G. R., Fisk, Z., Willis, J. O., and Smith, J. L. (1984). Possibility of Coexistence of Bulk Superconductivity and Spin Fluctuations in $UPt_3$. Phys. Rev. Lett. 52 679.

Stockert, O., Arndt, J., Faulhaber, E., Geibel, C., Jeevan, H., Kirchner, S. et al. (2011). Magnetically driven superconductivity in $CeCu_2Si_2$. Nat. Phys. 7, 119.

Sugitani, I., Okuda, Y., Shishido, H., Yamada, T., Thamizhavel, A., Yamamoto, E., et al. (2006). Pressure-Induced Heavy-Fermion Superconductivity in Antiferromagnet CeIrSi3 without Inversion Symmetry. J. Phys. Soc. Jpn. 75, 043703

Thomas, S. M., Santos, F. B., Christensen, M. H., Asaba, T., Ronning, F., Thompson, J. D., et al. (2020). Evidence for a pressure-induced antiferromagnetic quantum critical point in intermediate-valence $UTe_2$. Sci. Adv. 6, eabc870. URL http://advances.sciencemag.org/content/6/42/eabc8709.abstract.

Trovarelli, O., Geibel, C., Mederle, S., Langhammer, C., Grosche, F. M., Gegenwart, et al. (2000). $YbRh_2Si_2$: Pronounced Non-Fermi-Liquid Effects above a Low-Lying Magnetic Phase Transition. Phys. Rev. Lett. 85, 626.

Vieyra, H. A., Oeschler, N., Seiro, S., Jeevan, H. S., Geibel, C., Parker, D., and Steglich, F. 2011. Determination of Gap Symmetry from Angle-Dependent $H_{c2}$ Measurements on $CeCu_2Si_2$. Phys. Rev. Lett. 106 207001.

Vollhardt, D., and Wölfle, P. (1990). The Superfluid Phases of Helium3 (Taylor & Francis, London.

Volovik, G. E. (2003). The Universe in a Helium Droplet (Oxford University Press, NY).

Walker, I. R., Grosche, F. M., Freye, D. M., and Lonzarich, G. G. (1997). The normal and superconducting states of CeIn3 near the border of antiferromagnetic order. Physica C 282-7, 3003-306.

Wang, X.-P. , Qian, T., Richard, P., Zhang, P., Dong, J., Wang, H.-D., Dong, C.-H., Fang, M.-H. and Ding, H. (2011). Strong nodeless pairing on separate electron Fermi surface sheets in $(Tl, K)Fe_{1.78}Se_2$ probed by ARPES. Europhy. Lett. 93, 57001.

Weng, Z. F., Smidman, M., Jiao, L., Lu, X., and Yuan, H. Q. (2016). Multiple quantum phase transitions and superconductivity in Ce-based heavy fermions. Rep. Prog. Phys. 79, 094503.

White, B. D., Thompson, J. D., and Maple. M. B. (2015). Unconventional superconductivity in heavy-fermion compounds. J. Phys. C 514, 246-278. doi.org/10.1016/j.physc.2015.02.044.

Xu, M., Ge, Q. Q., Peng, R., Ye, Z. R., Jiang, J., Chen, F., Shen, X. P., Xie, B. P., Zhang, Y., Wang, A. F., Wang, X. F., Chen, X. H., and Feng, D. L. (2012). Evidence for an $s$-wave superconducting gap in $K_xFe_{2-y}Se_2$ from angle-resolved photoemission. Phys. Rev. B 85, 220504(R).

Xu, Y. J., Sheng, Y. T., and Yang, Y. F. (2019). Quasi-two dimensional fermi surfaces and unitary spin-triplet pairing in the heavy fermion superconductor $UTe_2$. Phys. Rev. Lett. 123, 217002.

Yamashita, .T, Takenaka, T., Tokiwa, Y., Wilcox, J. A., Mizukami, Y., Terazawa, D., Kasahara, Y., Kittaka, S., Sakakibara, T., Konczykowski, M., Seiro, S., Jeevan, H. S., Geibel, C., Putzke, C., Onishi, T., Ikeda, H., Carrington, A., Shibauchi, T., and Matsuda Y. (2017). Fully gapped superconductivity with no sign change in the prototypical heavyfermion $CeCu_2Si_2$. Sci. Adv. 3, e1601667.

Yoshida, T., Sigrist, M., and Yanase, Y. (2012). Pair-density wave states through spin-orbit coupling in multilayer superconductors. Phys. Rev. B 86, 134514.

Yuan, H. Q., Grosche, F. M., Deppe, M., Geibel, C., Sparn, G., Steglich, F. (2003). Observation of two distinct superconducting phases in $CeCu_2Si_2$. Science 302, 2104.

Zhai, H.-F., Jiao, W. H., Sun, Y. L., Bao, J. K., Jiang, H., Yang, X. J. et al. (2013). Superconductivity, charge- or spin-density wave, and metal–nonmetal transition in $BaTi_2(Sb_{1-x}Bi_x)_2O$. Phys. Rev. B 87, 100502.

Zhang, Z. Y., Chen, Z., Zhou, Y., Yuan, Y. F., Wang, S. Y., Wang, J. et al. (2021). Pressure-induced reemergence of superconductivity in the topological kagome metal $CsV_3Sb_5$. Phys. Rev. B 103, 224513.